\documentclass[pra,twocolumn,aps,showpacs,nofootinbib,floatfix,10pt]{revtex4-2}

\usepackage{amsmath, bm}
\usepackage{amssymb}
\usepackage{graphicx}
\usepackage{verbatim}
\usepackage[colorlinks=true,linkcolor=blue,citecolor=blue,urlcolor=blue]{hyperref}
\usepackage{txfonts}
\usepackage{dsfont}
\usepackage{xr}
\usepackage{enumitem}
\usepackage[utf8]{inputenc}
\usepackage{bbold}
\usepackage{mathtools}
\usepackage{lipsum, babel}
\usepackage{breqn}
\usepackage[normalem]{ulem}

\newcommand{\BHU}{Department of Physics, Institute of Science, Banaras Hindu University, Varanasi-221005, India}
\newcommand{\MPI}{Max Planck Institute for the Structure and Dynamics of Matter, Luruper Chaussee 149, 22761 Hamburg, Germany} 
\newcommand{\UH}{University of Hamburg, Luruper Chaussee 149, 22761 Hamburg, Germany} 
\newcommand{\HCUI}{The Hamburg Centre for Ultrafast Imaging, Luruper Chaussee 149, Hamburg D-22761, Germany}

\begin{document}

\title{Enhancing Measurement Precision of Non-Degenerate Two-Photon Absorption}

\author{Gaurav Shukla$^1$} \author{Shahram Panahiyan$^{2,3}$} \author{Devendra Kumar Mishra$^1$} \author{Frank Schlawin$^{2,3,4}$} \affiliation{$^1$ \BHU} \affiliation{$^2$ \MPI} \affiliation{$^3$ \UH} \affiliation{$^4$ \HCUI}

\begin{abstract}
Recent theoretical and experimental studies have shown that squeezed states of light can be engineered to enhance the resolution of nonlinear optical measurements. Here, we analyze non-degenerate two-photon absorption signals obtained from transmission measurements using two-mode squeezed light and compare different measurement strategies. In particular, we investigate how correlations between the light modes may be used to improve the
achievable precision. We find that intensity correlation measurements offer the best performance compared to normalized intensity correlation and noise reduction factor approaches. Under experimental imperfections modeled as linear photon losses, the enhancements from intensity and noise reduction measurements are reduced. In contrast, the normalized intensity correlation remains robust to loss, though this comes at the cost of losing the enhancement from non-classical light fields. This establishes a trade-off between robustness to loss and the achievable quantum advantage.
\end{abstract}

\maketitle

\section{\label{Introduction}Introduction}

Two-photon absorption (TPA) is a nonlinear optical process where a material is excited by the simultaneous absorption of two photons \cite{Mayer1931, RevModPhys.54.1061, PhysRevA.1.1696}
The experimental verification of TPA became feasible with the advent of the laser in 1961 \cite{PhysRevLett.7.229}, with the first experimental demonstration conducted in a europium-doped crystal \cite{PhysRevLett.7.229}. This breakthrough was soon followed by successful verification in other materials such as Caesium vapour \cite{PhysRevLett.9.453} and CdS \cite{PhysRev.134.A499}, underscoring the versatility and value of TPA across diverse scientific and technological domains. Due to its nonlinear nature, the TPA signal exhibits a quadratic dependence on the light intensity, in contrast to the linear dependence observed in single-photon absorption processes. TPA is further sensitive to the photon statistics and quantum correlations of the driving light fields. For instance, the absorption probability of entangled photon pairs scales linearly with the light field intensity \cite{Gea89, Javainen90}. This effect greatly facilitates nonlinear spectroscopy and microscopy at low photon fluxes \cite{Eshun2022, Dorfman2016, Schlawin2018, C6CS00442C, ATIF2007106}, offering significant advantages for the study of photosensitive samples. 

Recent developments have enabled efficient broadband time-domain TPA spectroscopy, particularly through single-shot methods using counter-propagating femtosecond pulses that avoid the need for temporal scanning and suppress nonlinear artifacts \cite{doi:10.1021/acs.analchem.4c01656}. The use of stimulated Raman scattering from the solvent as an internal reference further enhances measurement accuracy by accounting for intensity variations \cite{srivastava2023broadband}. Additionally, the strong light-matter interaction in semiconductor microcavities has been exploited to achieve high-speed optical switching via TPA \cite{donegan2002two, Maguire:06}.
Despite its significance, in different areas like in imaging \cite{PISTON199966, OHEIM2006788}, material science \cite{C8CS00849C, Sun:00}, quantum technologies \cite{Panahiyan2022, Panahiyan2023}, and medical applications \cite{C6CS00442C, ATIF2007106}, applications of TPA are restricted due to the small cross-sections resulting from weak light-matter coupling in free space. 

Traditional approaches employ strong laser pulses to enhance the excitation probability. However, this method is unsuitable for photo-sensitive materials such as biological samples. In this case, there has been an interest in leveraging the quantum properties of light to enhance nonlinear light-matter interactions \cite{Dorfman2016, Landes2024}. For example, techniques involving entangled photons \cite{Eshun2022, pandya2024, Lerch2021, Boitier2011, Dayan05, Dayan04} and squeezed states of light \cite{Carlos2021, Spasibko2017, Panahiyan2022} have shown to amplify nonlinear signals. Recent theory work further highlights opportunities arising from the use of nonclassical correlations and optimized measurements \cite{Dorfman2016, Schlawin2018, Yang2020, Zhang2022, PhysRevA.99.052318, Khan2024, Schaffrath2024, Schlawin2024, Karsa2024, Krstic2024}.
In our previous work \cite{Panahiyan2022}, we showed how the TPA measurement sensitivity may be enhanced with the use of optimally squeezed coherent states. Here, we aim to explore the possible improvement in the detection of the TPA signal using a two-mode squeezed state of light, produced by a non-degenerate optical parametric amplifier.
Our study aims to identify the optimal probe state for each correlation function, determine the necessary amount of seeding or squeezing, and identify the effects of the experimental imperfection in results.
In addition, we compare the performance of these optimal probe states to their classical counterparts  (TPA with two coherent light fields).

We consider three scenarios for the input light fields: squeezed vacuum, squeezed single-seeded coherent state, and squeezed double-seeded coherent state. We calculate correlation measurements of the transmitted fields and study the quantum correlation properties that contribute to reducing the uncertainty of the TPA signal. Specifically, we calculate the noise reduction factor \cite{Bondani07, Iskhakov16}, first-order correlation function (intensity correlation) \cite{gerry2023introductory, Loudon_book} and second-order correlation function (normalized intensity correlation) \cite{Loudon_book} and use them to calculate the error associated with measuring the TPA signal.
             
The paper is organized as follows: first, we establish the basis formalism and setup of our study in Section \ref{basics}. This includes introducing the model for the TPA process via a Markov master equation and squeezing process, modeling imperfect factors as single-photon loss, and introducing the correlation functions that we are interested in. Next in Section \ref{calculations}, we present a detailed study of optimizing the input light fields to attain the best 
estimation error
for the correlation functions and compare them as well as with their classical counterparts. In addition, we discuss our results from the phenomenological point of view and their interpretations.  Finally, Section \ref{Conclusion} provides a summary of our findings.

\section{Basics} \label{basics}
\subsection{Operator transformations}

\begin{figure*}
\centering
\includegraphics[width=0.9\linewidth]{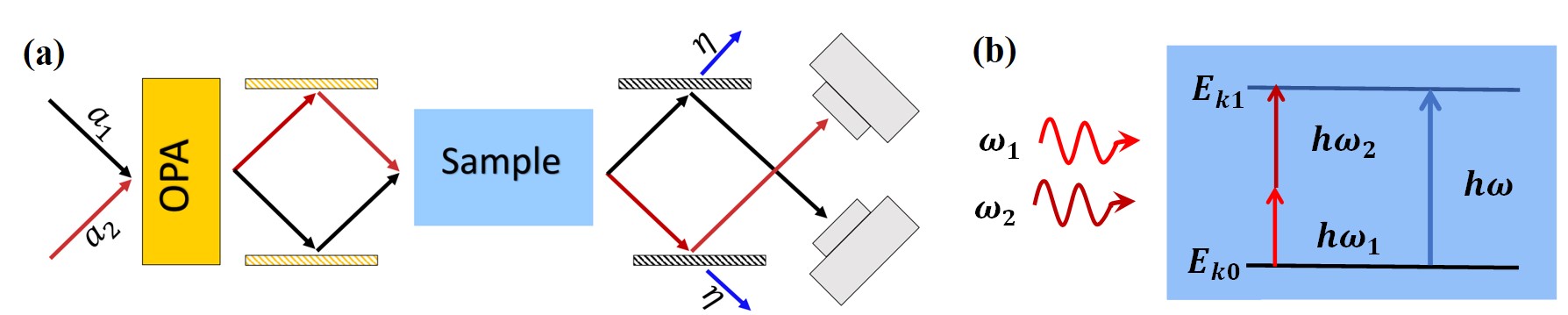}
\caption{Panel (a) the setup to calculate the measurement error of the TPA cross-section. Photon losses are modeled as imbalanced beam splitters with identical transmitivity $\eta$ in the setup. Panel (b) depicts the TPA process, where the material absorbs two photons simultaneously with frequencies $\omega_1$ and $\omega_2$, transitioning from the ground state, into an excited state, with frequency level $\omega$ in which $\omega_1 + \omega_2 = \omega$. }
\label{fig.1setup}
\end{figure*}

We consider the setup sketched in Fig. \ref{fig.1setup}(a): the two-mode squeezed light is incident on a sample and measurement is carried out on the transmitted light field.
Initially, we consider two modes of light characterized by annihilation operators $\hat{a}_1$ and $\hat{a}_2$. These two modes are prepared either in coherent or vacuum states. Denoting the density operator of the light field as $\hat{\rho}$, the optical parametric amplifier (OPA) is modeled by the transformation~\cite{BarnettRadmore} 
\begin{align}
\hat{\rho}_{r}=e^{ \hat{\mathcal{L}}_{OPA} }  \hat{\rho} &\equiv \hat{U}_{OPA } \hat{\rho} \hat{U}^{\dagger}_{OPA}. \label{eq.U_OPA}
\end{align}
Here, we have $\hat{U}_{OPA} = \exp \left[ \zeta \hat{a}^{\dagger}_1 \hat{a}^{\dagger}_2 - h. c. \right]$ \cite{BarnettRadmore, Lee90}, where $\zeta = r e^{i\theta}$ is the squeezing parameter with amplitude $r$ and phase $\theta$. 

Then, the squeezed state of light interacts with the sample. The TPA resonance in the sample is such that only the absorption of one photon from each mode can resonantly drive the excitation process, i.e. $\omega_1 + \omega_2 = \omega$, while $2\omega_1$ and $2\omega_2$ do not couple resonantly (see Fig. \ref{fig.1setup}(b)). We model this process as a Markovian open quantum system dynamics~\cite{Agarwal1970, Zubairy1980} considering that coupling between the system (sample) and bath (light field) is weak and our sample includes an ensemble of the systems that have two levels. Explicitly, the evolution equation reads
\begin{align}
\frac{d}{dt} \hat{\rho}_{r} &=\gamma_{TPA} \hat{\mathcal{L}}_{TPA} \hat{\rho}_{r}=
\notag 
\\
&= \frac{\gamma_{TPA}}{2} \left( 2 \hat{a}_1 \hat{a}_2 \hat{\rho}_{r} \hat{a}_2^\dagger \hat{a}_1^\dagger - \hat{a}_2^\dagger \hat{a}_1^\dagger \hat{a}_1 \hat{a}_2 \hat{\rho}_{r} - \hat{\rho}_{r} \hat{a}_2^\dagger \hat{a}_1^\dagger \hat{a}_1 \hat{a}_2\right), \label{eq.Lindblad}
\end{align}
where $\hat{\rho}_{r}$ is the density operator of the squeezed state and the Lindblad operator $\hat{L} = \hat{a}_1 \hat{a}_2$ describes the simultaneous loss of two photons. $\gamma_{TPA}$ is the absorption coefficient related to the absorbance via $\epsilon \equiv \gamma_{TPA} t$ with $t$ being the time duration of light propagating through the sample. The cross-section of the two-photon absorption process is a linear function of the absorption coefficient, and consequently the absorbance as well \cite{Panahiyan2023, Agarwal1970}. With Eq. (\ref{eq.Lindblad}), the light field after interaction with the sample is given by
\begin{align}
\hat{\rho}_{r,\epsilon}=e^{\epsilon \hat{\mathcal{L}}_{TPA} }\hat{\rho}_{r}.
\end{align}
Before the detection, a part of the light field can be lost due to different optical imperfections. We model these imperfections by considering a fictitious beam splitter having transmitivity $\eta$ ($\eta = 1$ corresponding to no loss while $\eta = 0$ indicates complete photon loss) on the outputs of the TPA (Fig. \ref{fig.1setup} (a)) \cite{BarnettRadmore, Drummond, Panahiyan2022}. This mathematically reads as
\begin{align} \label{eq.L_loss}
\hat{\rho}_{r,\epsilon,\eta}=e^{ \hat{\mathcal{L}}_{\eta }} \hat{\rho}_{r,\epsilon} &= \hat{U}_{\eta  } \hat{\rho}_{r,\epsilon} \hat{U}^{\dagger}_{\eta },
\end{align}
in which $\hat{U}_{\eta } = \exp \left(\tau \frac{\hat{a}_{i} \hat{u}_{i}^\dagger + \hat{u}_{i} \hat{a}_i^\dagger}{2} \right)$ where $\tau =  \arccos (\sqrt{\eta} )$ and $\hat{u}_i$ is a photon annihilation operator in an auxiliary mode that remains in the vacuum state. 

The expectation value of an observable $\hat{O}$ is given by 
\begin{align}
\langle \hat{O} \rangle &= \text{tr} \left\{ \hat{O} e^{\hat{\mathcal{L}}_{\eta}}  e^{\epsilon \hat{\mathcal{L}}_{TPA}} e^{ \hat{\mathcal{L}}_{OPA} } \hat{\rho}' \right\}, \label{eq.General}
\end{align}
where $\hat{\rho}' = \hat{\rho}_{r,\epsilon,\eta} \otimes |0\rangle\langle 0|$ with photon loss bath density operator $|0\rangle\langle 0|$. In Eq. \eqref{eq.General}, we consider $\epsilon$ to be very small, i.e., $\epsilon \ll 1$. This assumption is justified in most applications since the two-photon absorption cross-sections are typically very small \cite{Boyd}. This allows us to make the approximation $e^{\epsilon \hat{\mathcal{L}}_{TPA}} \simeq \mathbb{1} + \epsilon \hat{\mathcal{L}}_{TPA}$ and the transmitted density matrix can be expressed as $\rho_{r,\epsilon} \simeq \rho_{r} + \epsilon (\partial \rho_{r}/ \partial \epsilon)$. 

By using the property of the trace, operators in the parenthesis can be rewritten and defined as 
\begin{equation}
 \langle \hat{O} \rangle  = \text{tr} \left\{e^{\hat{\mathcal{L}}_{\eta}} (\mathbb{1} + \epsilon\hat{\mathcal{L'}}_{TPA}) e^{ \hat{\mathcal{L}}_{OPA} } \hat{O} \hat{\rho}' \right\} \label{eq.newG},
\end{equation}
where $\hat{\mathcal{L'}}_{TPA}$ is the adjoint of the superoperator given in Eq. (\ref{eq.Lindblad}) and acts on some operator $\hat{O}$ as 
\begin{equation}  \hat{\mathcal{L'}}_{TPA} \hat{O} = \frac{1}{2} \left( 2  \hat{a}_2^\dagger \hat{a}_1^\dagger \hat{O} \hat{a}_1 \hat{a}_2  - \hat{O} \hat{a}_2^\dagger \hat{a}_1^\dagger \hat{a}_1 \hat{a}_2  - \hat{a}_2^\dagger \hat{a}_1^\dagger \hat{a}_1 \hat{a}_2 \hat{O}\right). \label{adj.Lindblad}
\end{equation}
Using Eqs. (\ref{eq.U_OPA})$-$(\ref{adj.Lindblad}), we can model the setup in Fig.~\ref{fig.1setup}(a) as successive steps of operator transformations, which let us evaluate observables most conveniently.

An initial two-mode coherent state is described by the displacement 
\begin{equation}
\begin{split}
    \hat{a}_1 = \hat{v}_1 + \alpha_1 e^{i\phi_1}, \\
\hat{a}_2 = \hat{v}_2 + \alpha_2 e^{i\phi_2},
\end{split}
\end{equation}
where $\hat{v}_1$ and $\hat{v}_2$ are the corresponding vacuum annihilation operators. $\alpha_{1,2}$ and $\phi_{1,2}$ are the coherent state's real positive amplitudes and phases. Next, the squeezing transformation is given by
\begin{equation}
\begin{split}
\hat{b}_1 = \cosh{r}~ \hat{a}_1 + e^{i \theta} \sinh{r}~ \hat{a}_2^\dagger, \\
\hat{b}_2 = \cosh{r}~ \hat{a}_2 + e^{i \theta} \sinh{r}~ \hat{a}_1^\dagger,
\end{split}
\end{equation}
with the parameters given below Eq.~(\ref{eq.U_OPA}). 
The interaction with the TPA sample results in the following output relations for the operators, using Eq. (\ref{adj.Lindblad}), 
\begin{equation}
\begin{split}
\hat{c}_1 = \hat{b}_1 - \frac{\epsilon}{2} \left(\hat{b}_2^\dagger \hat{b}_2 \hat{b}_1\right), \\
\hat{c}_2 = \hat{b}_2 - \frac{\epsilon}{2} \left(\hat{b}_1^\dagger \hat{b}_1 \hat{b}_2 \right).
\end{split}
\end{equation}
Note that the operators $ c_1 $ and $ c_2 $ do not satisfy the standard commutation relation $ [c, c^\dagger] = 1 $. This deviation arises due to the omission of terms involving higher-order powers of $\epsilon$.
Finally, the single-photon losses lead to 
\begin{equation}
\begin{split}
\hat{d}_1 = \sqrt{\eta} \hat{c}_1 + \sqrt{1 - \eta} \hat{u}_1, \\
\hat{d}_2 = \sqrt{\eta} \hat{c}_2 + \sqrt{1 - \eta} \hat{u}_2. \label{outoperator}
\end{split}
\end{equation}
Eq. (\ref{outoperator}) establishes the relation between the input and output operators.

\subsection{Correlation functions} 

Here, we investigate the precision of measurement of the TPA absorbance $\epsilon$ using correlation functions as observables. The uncertainty in the measurement 
using an operator, $\hat{O}$, is written based on error propagation
(also known as the method of moments in quantum metrology \cite{Pezz2018})
as
\begin{equation}
    \Delta \epsilon^2_{\hat{O}} = \frac{ \text{Var} (\hat{O}) }{ \left( \frac{\partial \langle \hat{O} \rangle}{\partial \epsilon} \right)^2_{\epsilon=0} }, \label{generalError}
\end{equation}
where $\hat{O}$ in this equation is an operator having the uncertainty Var($\hat{O}$) depending on the quantum state $\hat{\rho}_{r,\epsilon,\eta}$
(for more details, please see Appendix \ref{appendix error}) and we evaluate it at the limit of $\epsilon \rightarrow 0$. From Eq. (\ref{eq.newG}), variation in expectation value of $\hat{O}$ with respect to $\epsilon$ can be written as
\begin{equation}
   \left. \frac{\partial \langle \hat{O} \rangle}{\partial \epsilon}\right|_{\epsilon = 0} = \text{tr} \left\{ e^{\hat{\mathcal{L}}_{\eta}} \hat{\mathcal{L'}}_{TPA} e^{ \hat{\mathcal{L}}_{OPA} } \hat{O} \  \hat{\rho}' \right\}. \label{eq7}
\end{equation}

We evaluate the intensity measurements in both output ports using the above equations and calculate the following correlation functions:

\textit{Noise reduction factor}:
The noise reduction factor (NRF) is defined as \cite{Bondani07, Iskhakov16}
\begin{align}
\mathcal{N} &\equiv \frac{ \text{Var} (\hat{n}_1 - \hat{n}_2) }{\langle \hat{n}_1 \rangle + \langle \hat{n}_2 \rangle}, \label{eq.NRF}
\end{align}
where $\langle \hat{n}_1 \rangle$ and $\langle \hat{n}_2 \rangle$ are the mean photon counts in the output modes. If the variance of the difference $\text{Var} (\hat{n}_1 - \hat{n}_2) < \langle \hat{n}_1 \rangle + \langle \hat{n}_2 \rangle$,
it indicates quantum correlations between the output modes \cite{Bondani07, Iskhakov16}. In contrast, for $\text{Var} (\hat{n}_1 - \hat{n}_2) \geq \langle \hat{n}_1 \rangle + \langle \hat{n}_2 \rangle$, the output modes possess classical correlations  \cite{Bondani07, Iskhakov16}. We obtain the error in the case of NRF measurement using Eq. \eqref{generalError} in which $\text{Var}(\mathcal{N})$ and $({\partial \mathcal{N}}/{\partial \epsilon})$ are calculated by using Eqs. (\ref{NRFpartial}) and (\ref{NRFvariance}).

\textit{First-order intensity correlation function $G(1,1)$}:
The significance of the first-order intensity correlation function in quantum optics lies in its relationship with the intensity. It provides valuable insights into the statistical properties of light and facilitates the study of quantum interference and correlations. Here, we consider the cross-correlation between the two output ports, given by
the product of photon number operators $\hat{n}_1$ and $\hat{n}_2$ of the two output modes \cite{gerry2023introductory, Loudon_book}
\begin{equation}
    G(1,1) = \langle \hat{n}_1 \hat{n}_2 \rangle.
\end{equation}
From Eq. (\ref{generalError}), we find the error in the measurement of the TPA which we will use to evaluate precision accessible via $G(1,1)$ measurement. It should be noted that this quantity gives us the intensity correlation between the modes. 

\textit{Second-order correlation function $g(1,1)$}: 
The second-order correlation function $g(1,1)$ is the ratio of the probability of detecting photon numbers on two ports to the product of the probability of detecting photon numbers at each of these. It is given by \cite{Loudon_book}
\begin{align}
g(1,1) &\equiv \frac{ \langle \hat{n}_1 \hat{n}_2 \rangle }{\langle \hat{n}_1 \rangle  \langle \hat{n}_2 \rangle}. \label{eq.g2}
\end{align}

$g(1,1)$ provides valuable insights into the nonclassical statistics of light  \cite{PhysRevA.42.1608, gerry2023introductory, Loudon_book}: 
$g(1,1) \geq 1$ indicates photon bunching with potentially classical correlations \cite{PhysRevA.42.1608}. 
In contrast, $g(1,1) < 1$ indicates photon anti-bunching, which is indicative of the nonclassical nature of light \cite{PhysRevA.90.063824}. We again follow Eq. (\ref{generalError}) to find $\Delta \epsilon_{g(1,1)}^2$ where $\text{Var} (g(1,1))$ and $({\partial g(1,1)}/{\partial \epsilon})$ are calculated using Eqs. (\ref{g2partial}) and (\ref{g2variance}). It should be noted that this quantity gives us the normalized intensity correlation between the modes.

\section{Estimation error with classical and squeezed light fields}\label{calculations}
Using the three correlation functions mentioned above, we calculate the error for the measurement of the TPA coefficient, $\Delta\epsilon^2$, namely estimation error. We consider three scenarios for transmission measurements with 
(i) a two-mode squeezed vacuum, 
(ii) a squeezed state, where a coherent state seeds one input mode, and 
(iii) a two-mode squeezed state, where both modes are seeded with coherent states. After presenting our results, we will compare and discuss implications in section \ref{Discussion}.
The scaling behaviours are studied for large photon numbers ($\geq 100$).

\subsection{Coherent light fields}

To establish a baseline for how much squeezed light can improve the precision of the TPA measurements, we first study the classical case, i.e., the precision for measuring the TPA absorbance with two coherent states of light. Using the formalism of section \ref{basics}, we find the scaling behaviour of $\Delta \epsilon^2_{NRF}$  
\begin{equation}
    \Delta \epsilon^2_{NRF_{Classical}} \approx \frac{8}{n_T^2 \eta^2},
\end{equation}
where $n_T$ is the total photon number reaching the sample.
For $G(1,1)$ measurements, we find
\begin{equation}
    \Delta \epsilon^2_{G(1,1)_{Classical}} \approx \frac{4}{n_T^{3} \eta},
\end{equation}
and for $g(1,1)$
\begin{equation}
    \Delta \epsilon^2_{g(1,1)_{Classical}} \approx \frac{4}{n_T^2 \eta^2}.
\end{equation}
Hence, at large photon numbers, measurements of $G(1,1)$ outperform the other correlation measurements, providing both an improved cubic scaling of the $\Delta \epsilon^2$ and a reduced sensitivity to single-photon losses. The worst precision may be extracted from the NRF.

\subsection{Squeezed vacuum}

The NRF with a two-mode squeezed vacuum evaluates to 
\begin{equation}
    \mathcal{N} = (1 - \eta).
\end{equation}
Crucially, it does not depend on the TPA absorbance $\epsilon$. Hence, NRF does not capture any information about the TPA losses. 
It only reflects the amount of single-photon loss taking place in the setup which degrades the amount of correlations between the two beams and thereby increases the NRF. 

\begin{figure}
\centering
\includegraphics[width=0.3\textwidth]{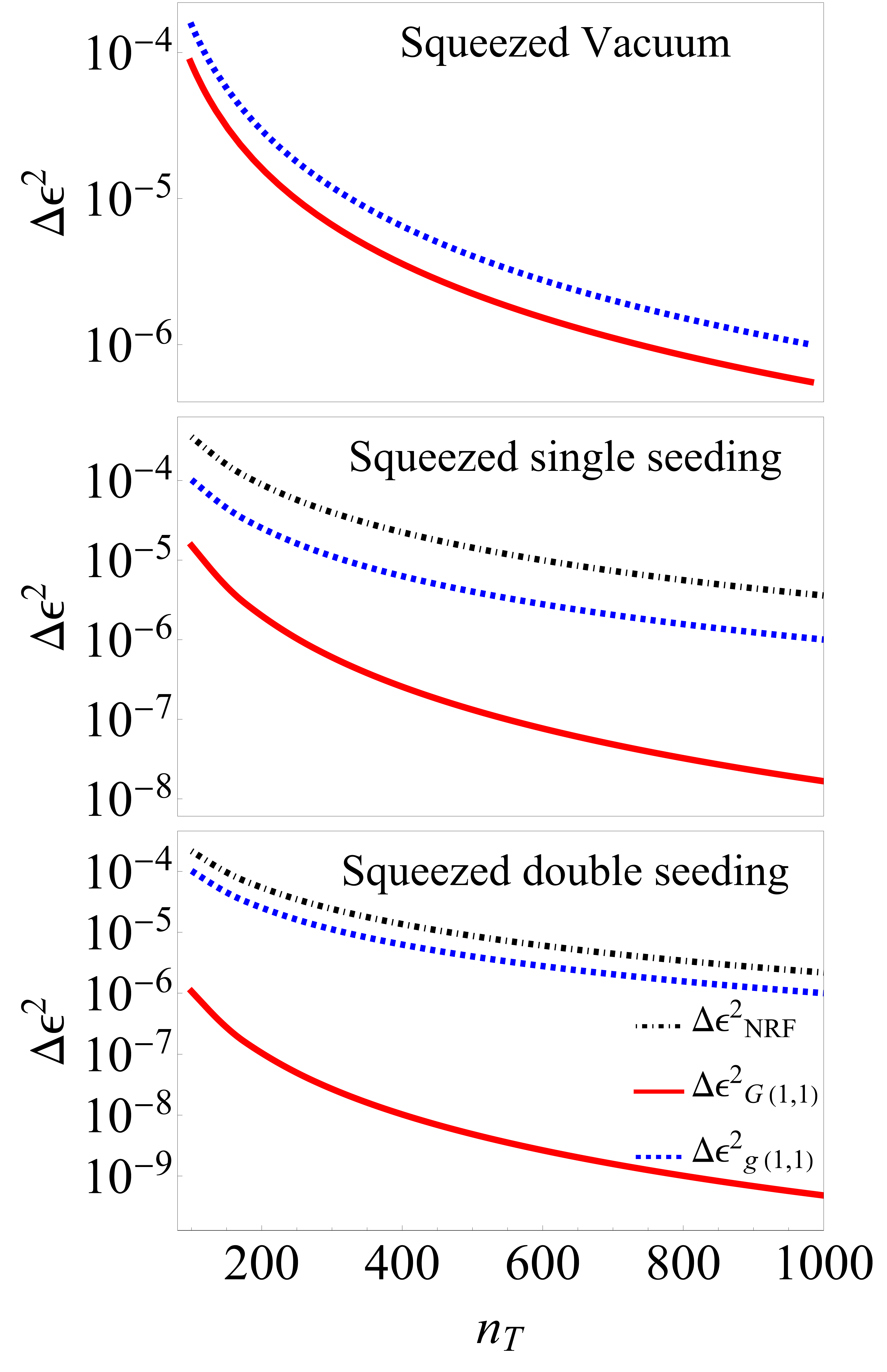}
\caption{ 
Optimum $\Delta \epsilon^2$ as a function of the total photon number for three cases of the squeezed vacuum (upper panel), the squeezed single seeding (middle panel), and the squeezed double seeding (lower panel) initial state of light. In the absence of seeding, the NRF is insensitive to two-photon absorption, therefore it is absent in the upper panel. Irrespective of the type of the state of light interacting with the sample, measurements of the correlation function $G(1,1)$ provide the best precision for TPA absorbance measurements.}
\label{errorVSnTthreeCa}
\end{figure}

\begin{figure*}
\centering
\includegraphics[width=0.32\linewidth]{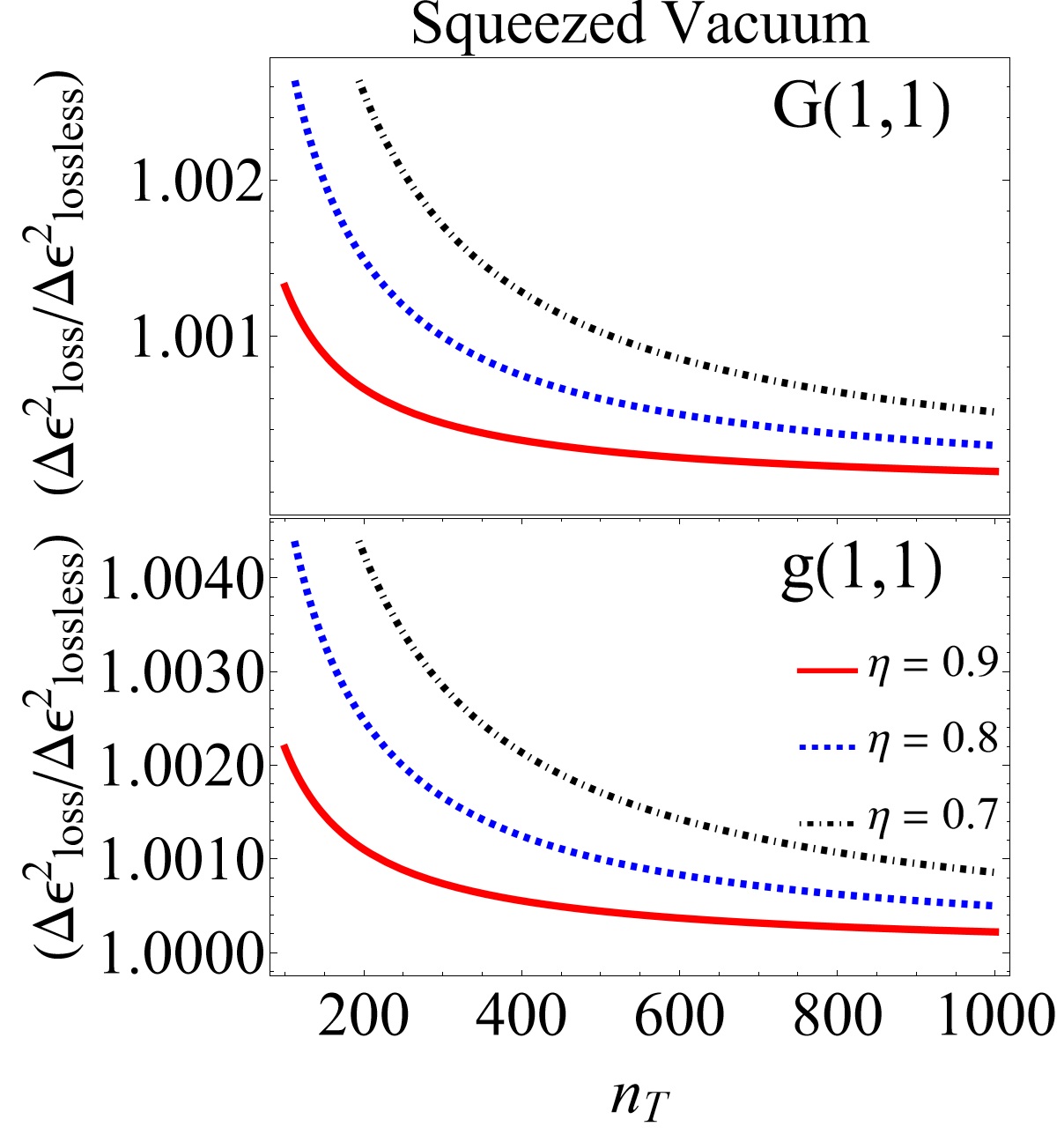}
\includegraphics[width=0.335\linewidth]{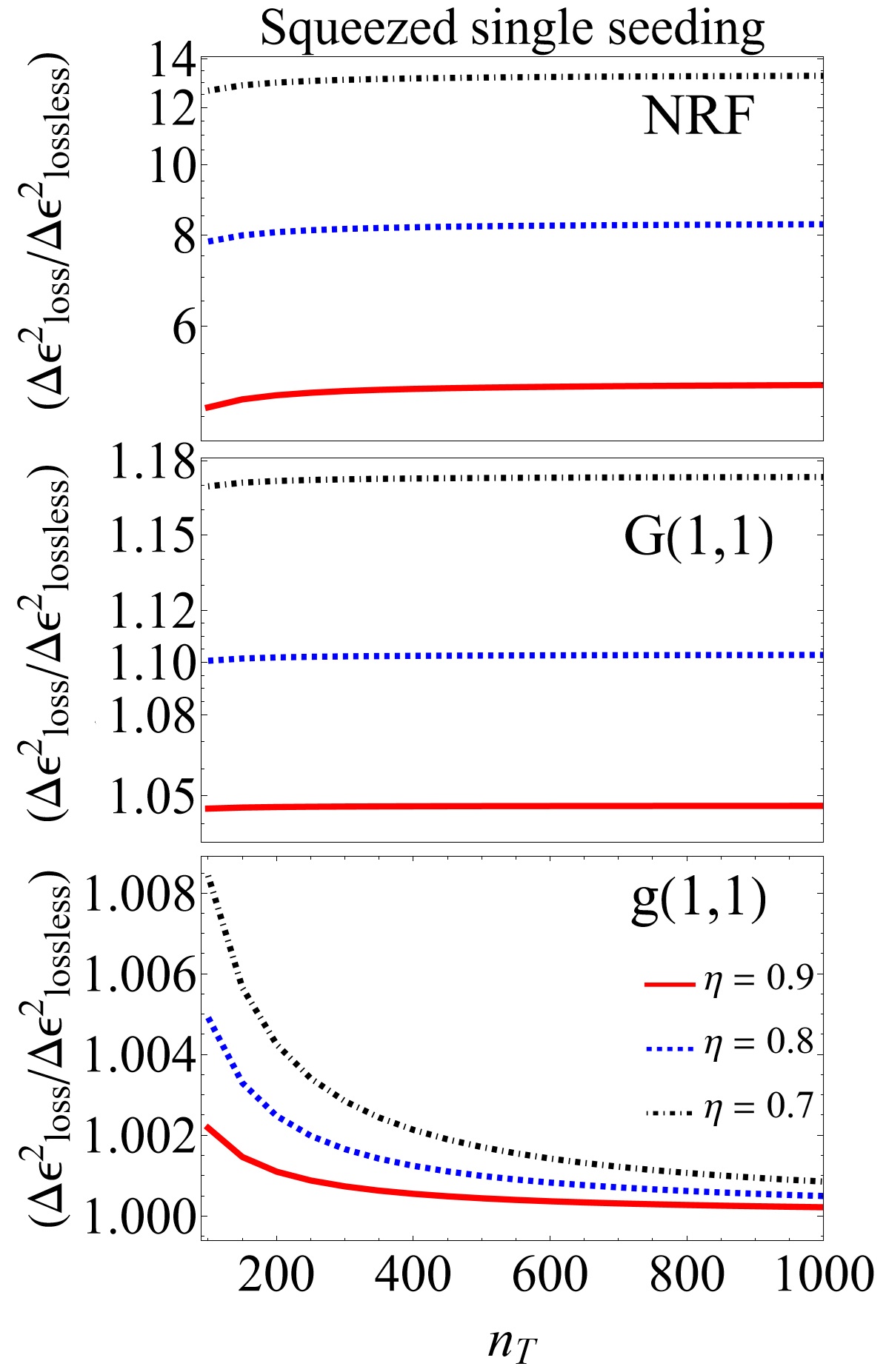}
\includegraphics[width=0.335\linewidth]{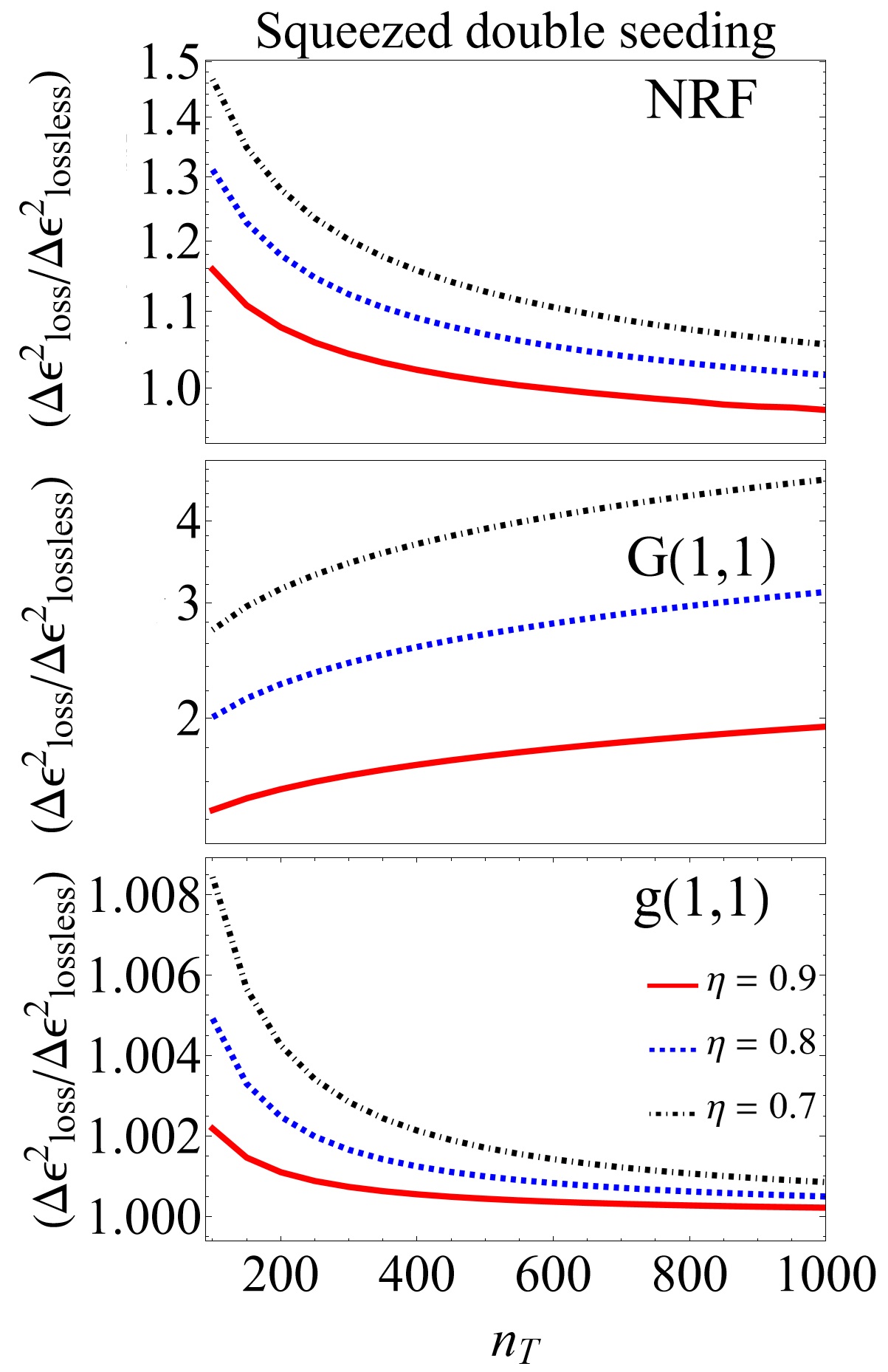}
\caption{ 
The normalized error, $\Delta \epsilon^2_{\text{loss}}/\Delta \epsilon^2_{\text{lossless}}$, 
for squeezed vacuum (left panels), optimized squeezed single-seeded (middle panels) and squeezed double-seeded states (right panels). We set the relative phase for the right panels according to Fig. \ref{DSphaseNRFG11g2lossless}. For both the middle and right panels, the optimized squeezing parameter for different total photon numbers is given in Fig. \ref{SSnTrForNRFG11g2}.} 
\label{NormalisedError}
\end{figure*}

The  $\Delta \epsilon^2$ using $G(1,1)$ measurements yields
\begin{equation}
    \Delta \epsilon_{G(1,1)}^2 = \frac{2 + 5 n_T^3 \eta^2 + 2 n_T^2 \eta (3 + 5 \eta) + n_T (2 + 8 \eta + 3 \eta^2)}{n_T (1 + 5 n_T + 3 n_T^2)^2 \eta^2},\label{G11ErrorTMS}
\end{equation}
where $n_T = 2 \sinh^2{r}$ is the total photon number $\langle \hat{n}_1 + \hat{n}_2 \rangle$ incident on the TPA sample. 
From Eq. (\ref{G11ErrorTMS}), we can see that  $\Delta \epsilon^2$ is independent of the OPA phase parameter $\theta$. 
At large photon numbers, Eq. (\ref{G11ErrorTMS}) scales as
\begin{equation} 
    \Delta \epsilon^2_{G(1,1)} \approx \frac{5}{9 n_T^2}.
\end{equation}
In this limit, the estimation error $\Delta \epsilon^2$ becomes proportional to the square of the total photon number. As shown in the upper panel of Fig. \ref{errorVSnTthreeCa}. 
In addition, it becomes independent of single-photon losses in the large photon number regime which can be observed in the left panels of Fig. \ref{NormalisedError}.  

In case of the $g(1,1)$, the estimation error $\Delta \epsilon^2$ reads
\begin{equation}
    \Delta \epsilon_{g(1,1)}^2 = \frac{2 - 4 \eta + 4 \eta^2 + n_T^3 \eta^2 + 2 n_T^2 \eta (1 + \eta) + n_T (2 + 3 \eta^2)}{n_T \eta^2 (-1 + n_T + n_T^2)^2},
\end{equation}
and at large photon numbers, the scaling behavior is given by
\begin{equation}
    \Delta \epsilon^2_{g(1,1)} \approx \frac{1}{n_T^2}.
\end{equation}
Similar to the $G(1,1)$ scenario, the estimation error $\Delta \epsilon^2$ scales quadratically with the total photon number, as illustrated in the upper panel of Fig. \ref{errorVSnTthreeCa}, and becomes independent of single-photon losses at large photon numbers as shown in Fig. \ref{NormalisedError}. It should be noted that $\Delta \epsilon^2_{g(1,1)}$ is independent of the squeezing phase.

\begin{figure}
\centering
\includegraphics[width=0.45\textwidth]{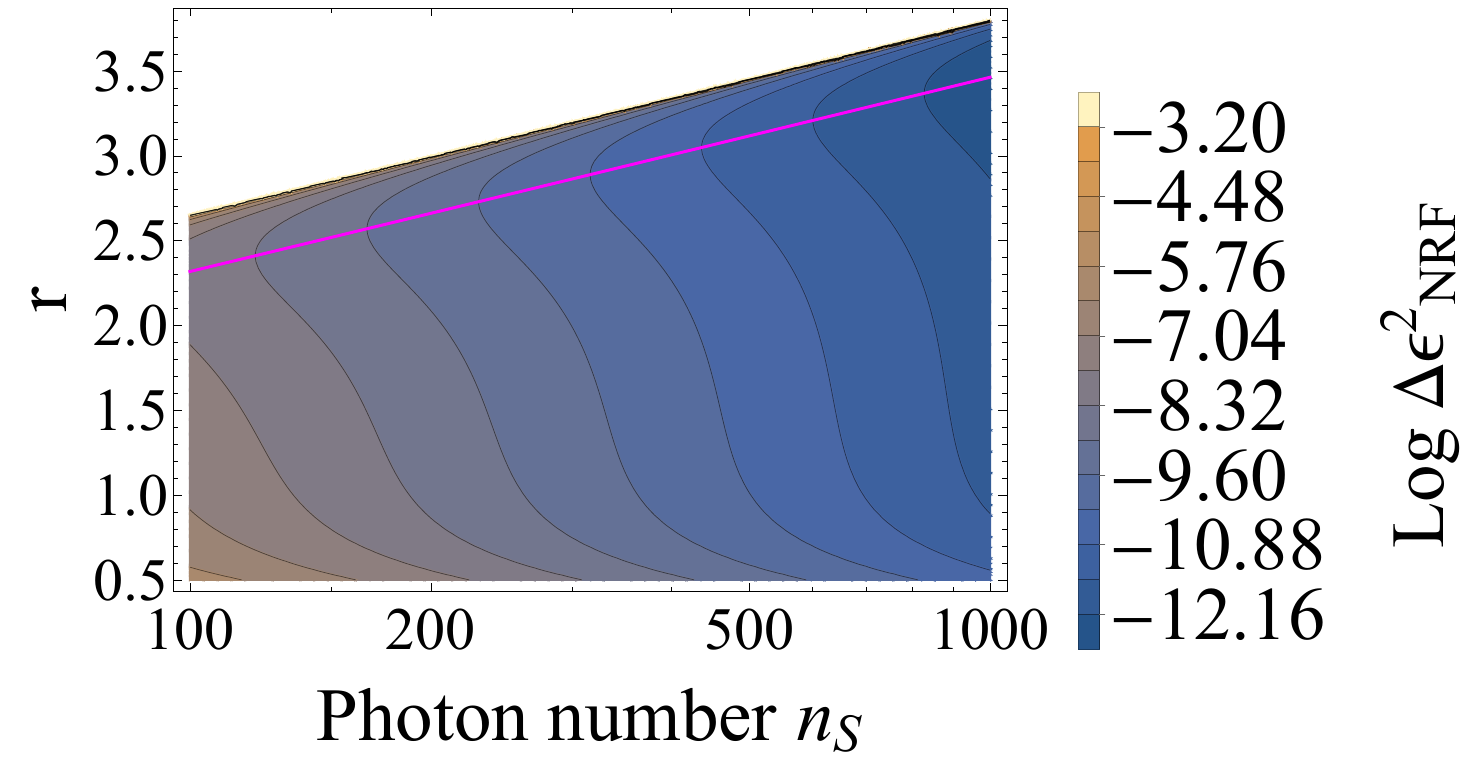}
\caption{ 
Estimation error $\Delta \epsilon^2$ as a function of the total photon number, $n_T$, and squeezing parameter, $r$, for the lossless case is shown. The magenta line maps optimal regimes of squeezing parameters for different total photon numbers. 
}
\label{CaseEx}
\end{figure}

\begin{figure*}
\centering
\includegraphics[width=0.38\linewidth]{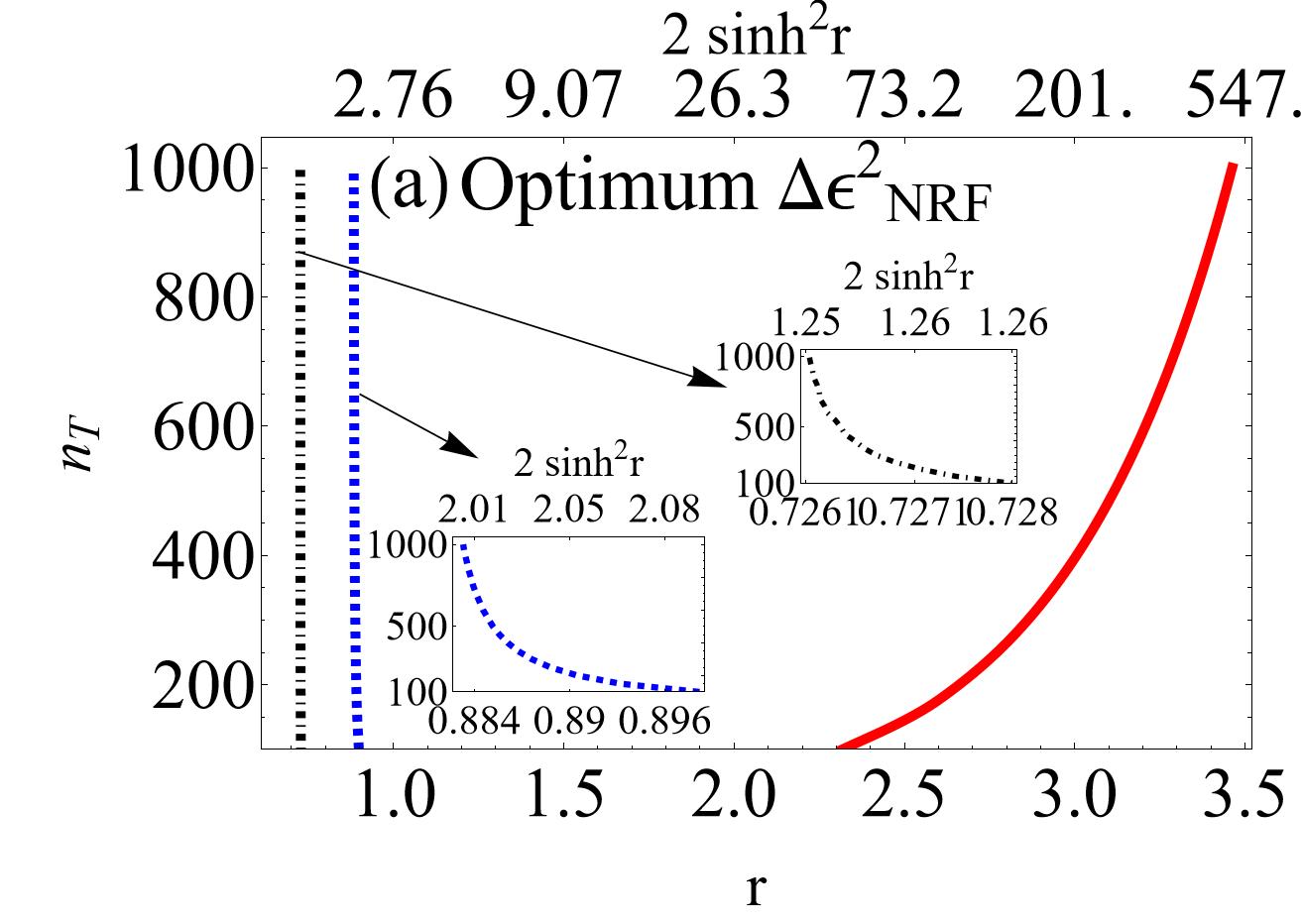}
\includegraphics[width=0.30\linewidth]{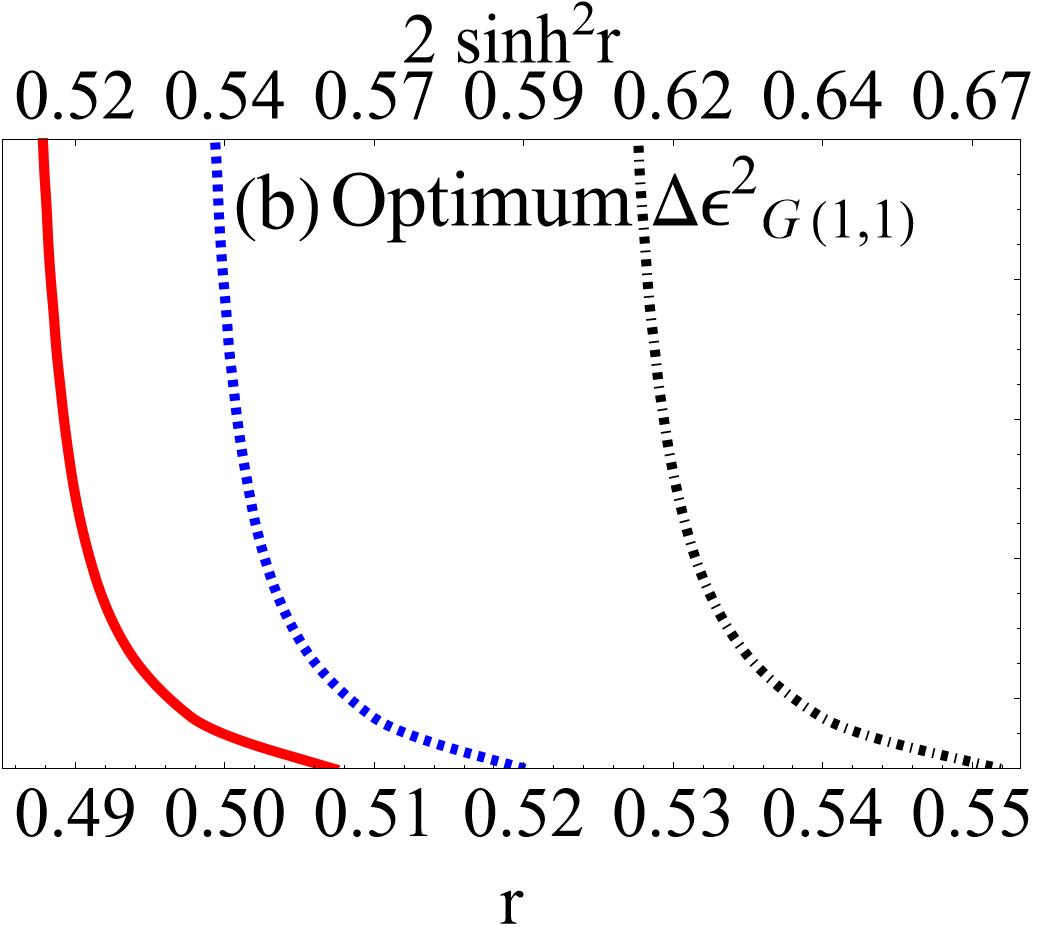}
\includegraphics[width=0.307\linewidth]{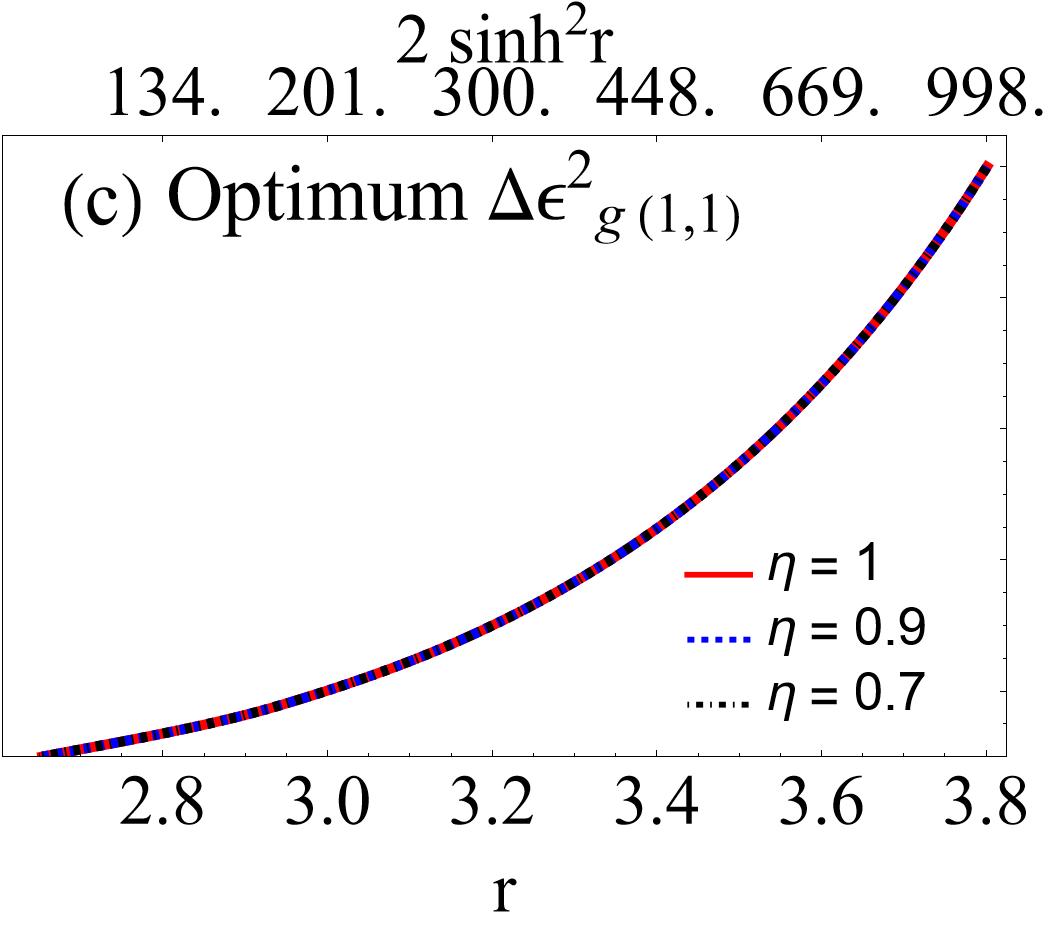}
\includegraphics[width=0.37\linewidth]{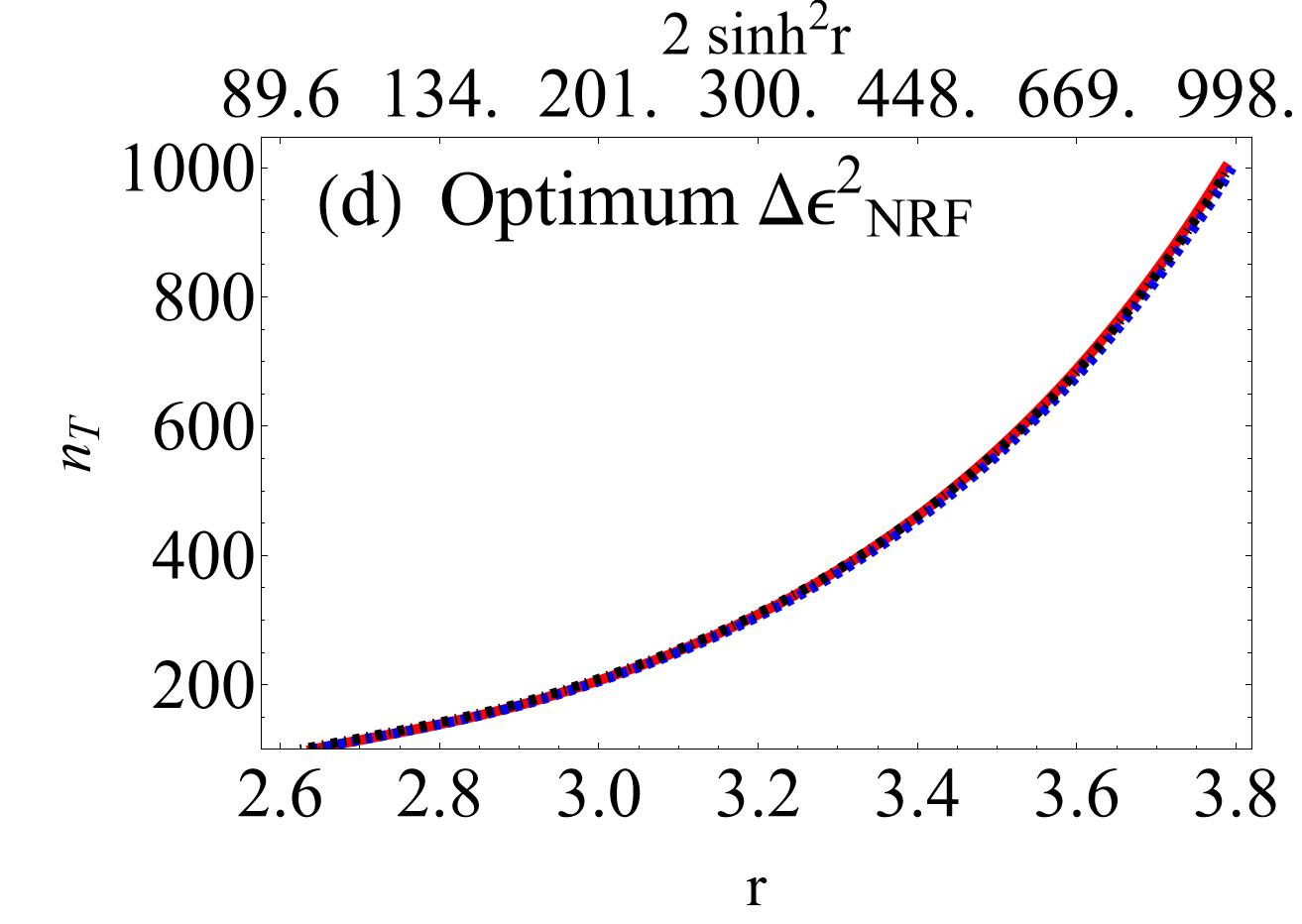} \includegraphics[width=0.305\linewidth]{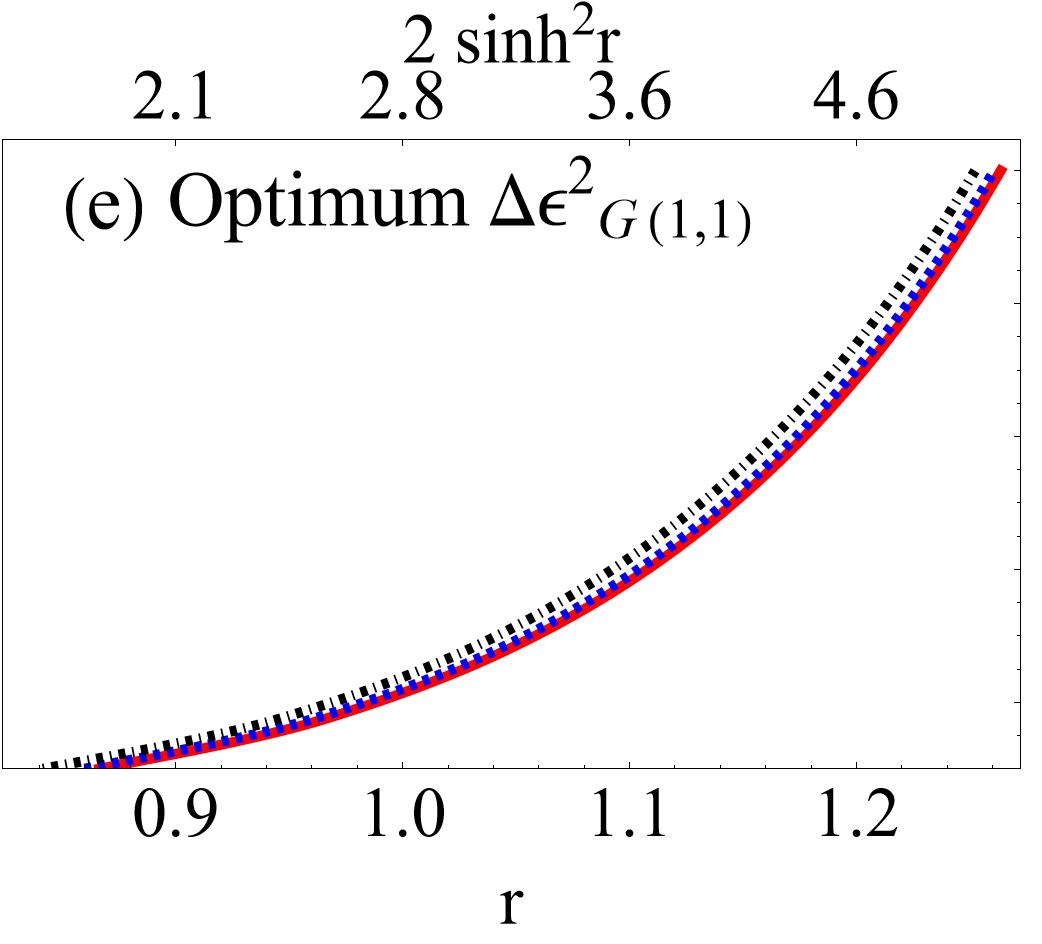}
\includegraphics[width=0.311\linewidth]{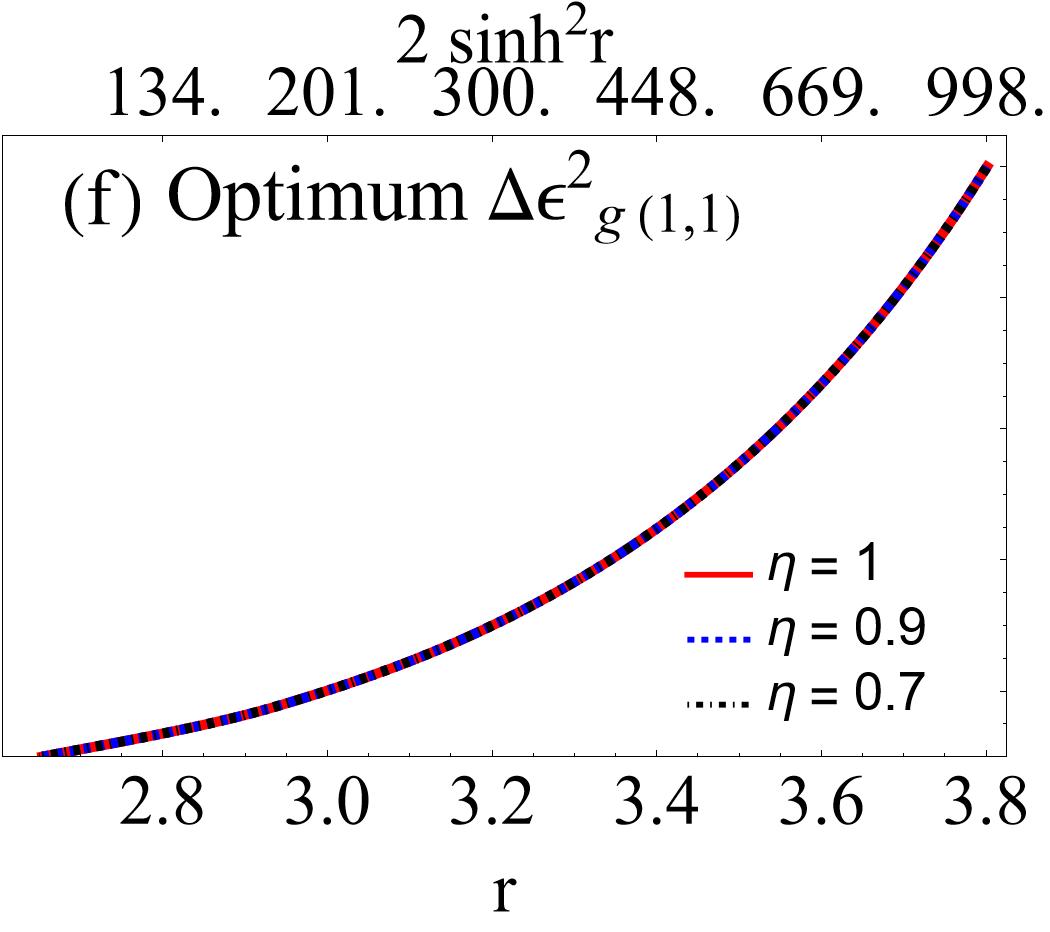}
\caption{Variation of $n_T$ versus $r$ in case of optimal $\Delta \epsilon^2$ with single seeding (upper panels) and double seeding (lower panels) for NRF (left panels), $G(1,1)$ (middle panels), and $g(1,1)$ (right panels). For the lower panels, we set the relative phase according to Fig. \ref{DSphaseNRFG11g2lossless}.}
 \label{SSnTrForNRFG11g2}
\end{figure*}

\begin{figure*}
\centering
\includegraphics[width=1\textwidth]{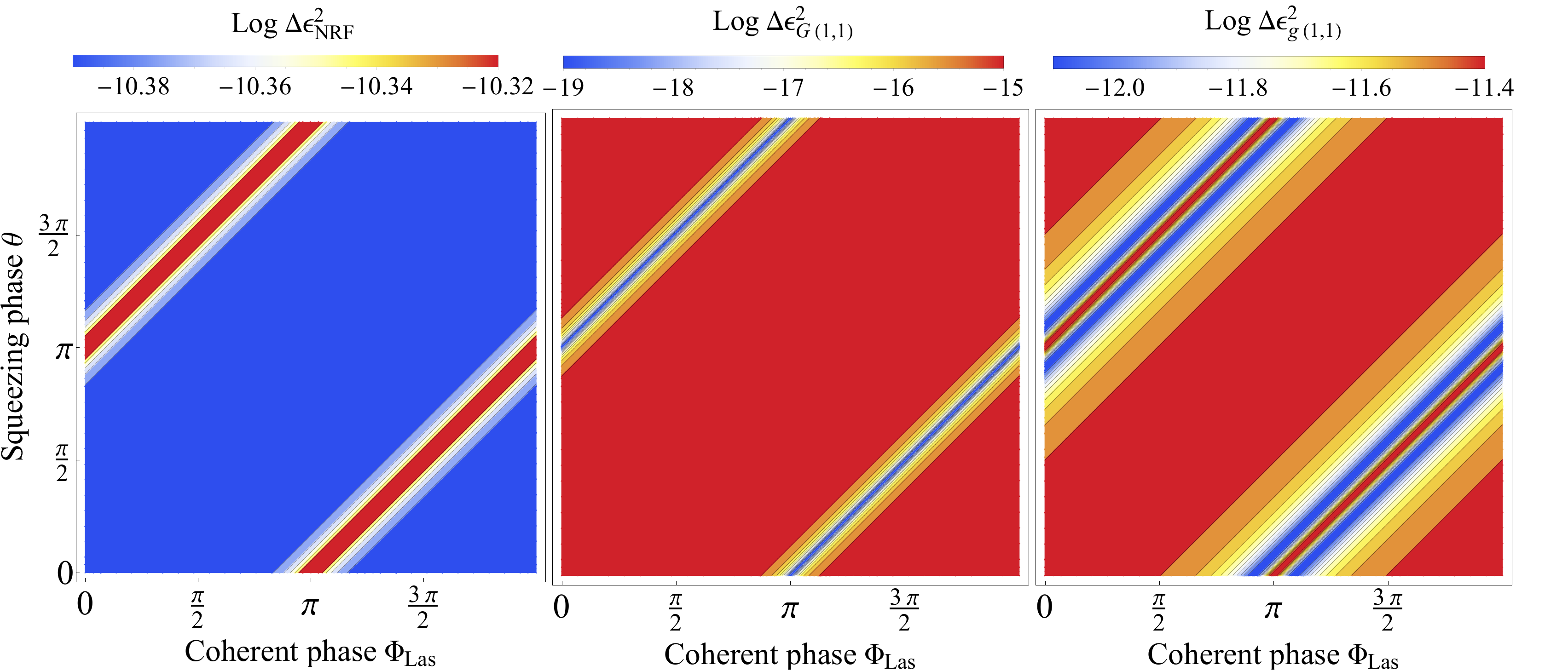}
\caption{
Phase dependence of the $\Delta \epsilon^2$ as a function of $\theta$ and relative phase $\Phi$ for (a) $\Delta \epsilon^2_{NRF}$, (b) $\Delta \epsilon^2_{G(1,1)}$ and (c) $\Delta \epsilon^2_{g(1,1)}$ in double seeded cases with $n_T=500$, $r=1$, and $\eta=1$. We observe that optimum relative phase for NRF, $G(1,1)$, and $g(1,1)$ are $\theta-\Phi = 0$,and $2\pi$, $\theta-\Phi = \pm \pi$ with $j \in \mathbb{Z}$, and $\theta-\Phi \approx 10\pi/11$ and $12\pi/11$, respectively.}
\label{DSphaseNRFG11g2lossless}
\end{figure*}

\begin{figure}
\centering
\includegraphics[width=0.315\textwidth]{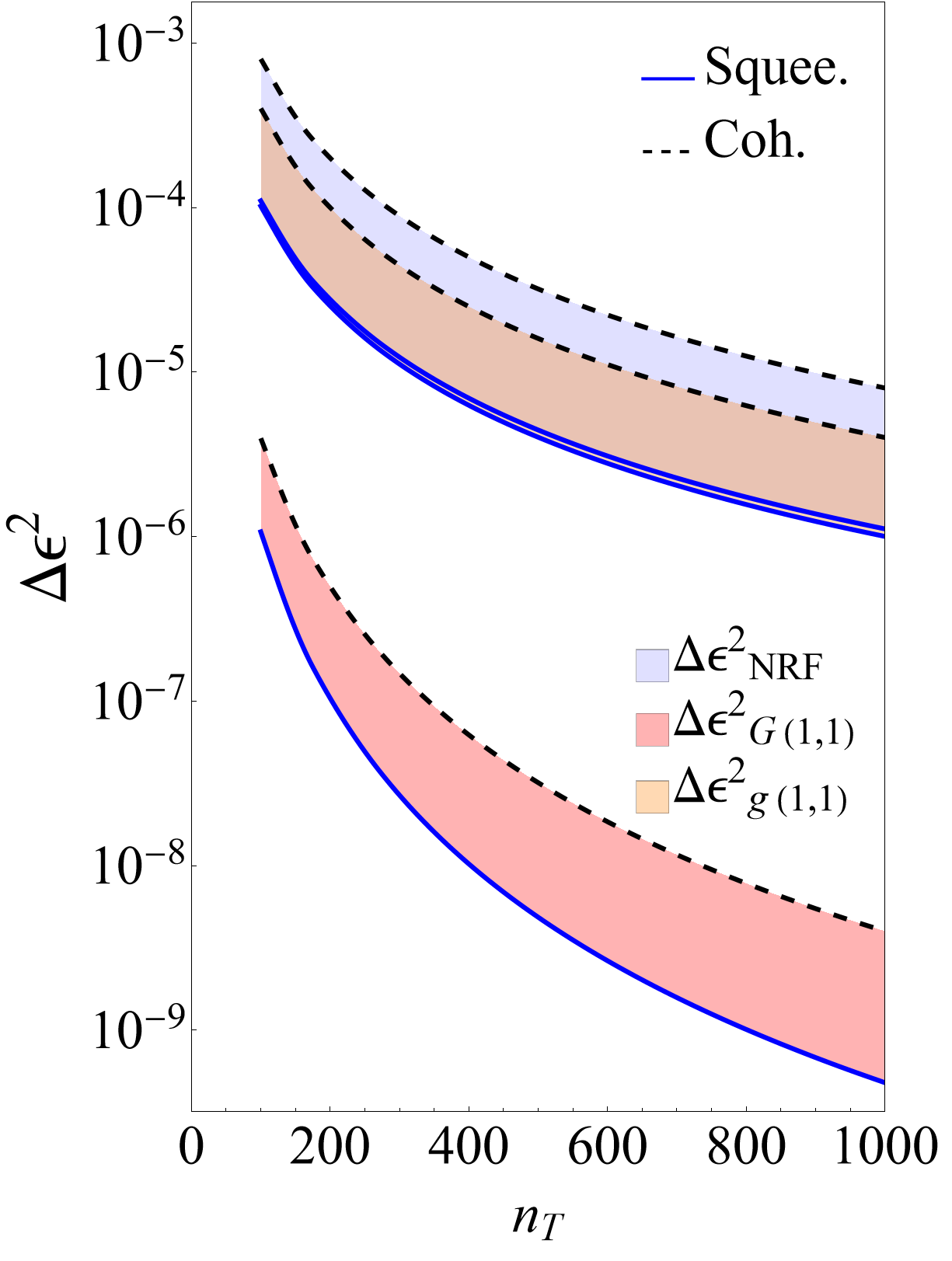}
\caption{ 
Optimum $\Delta \epsilon^2$ as a function of the mean photon number for two-mode coherent (dashed lines) and squeezed coherent (solid lines) light fields. We observe enhancements in the precision of absorbance for squeezed light fields compared to their classical counterparts for different correlation functions. The best precision to measure absorbance is provided by $G(1,1)$ measurement with the squeezed light field while the worst is provided by NRF with classical light fields.}
\label{ClassVSQuant}
\end{figure}

\subsection{Squeezed single-seeded coherent state}

In this scenario, we seed a coherent state (with amplitude $\alpha_1$ and phase $\phi_1$) in one of the input ports of the OPA and drive the sample with the resulting squeezed coherent state of light and vacuum-squeezed light from the other port. We fix the total photon number at the sample and optimize the squeezing parameter depending on the single-photon loss to obtain the best 
estimation error
for different correlation functions. It means we rewrite the seeding amplitude $\alpha_1$ as a function of the other parameters, the total photon number $n_T$, and the squeezing parameter $r$. The resulting optimization landscape is shown in Fig. \ref{CaseEx} for the lossless case of NRF. Repeating the same for different levels of single photon loss and other correlation functions gives us Fig. \ref{SSnTrForNRFG11g2}. 
In all three scenarios, the estimation error $\Delta \epsilon^2$ is independent of the relative phase between the seeding ($\phi_1$) and the OPA ($\theta$) phases. 
Next, we study the performance of the optimal 
estimation error
for these correlation functions.  

We find the estimation error $\Delta \epsilon^2$ for measurements of the NRF using the formalism of section \ref{basics}. The resulting extensive expressions are provided in a companion Mathematica file for brevity. 
The first significant effect of seeding is that the NRF becomes sensitive to the TPA process. In other words, seeding may be used to switch TPA detection on and off
and thus distinguish single- from two-photon losses. 
In the absence of single photon losses, the scaling behaviour of optimized $\Delta \epsilon_{NRF}^2$ is approximately given by
\begin{equation}
    \Delta \epsilon_{NRF}^2 \approx \frac{3}{n_T^2}.
\end{equation} 
Here, the total photon number is given by $n_T = \alpha_1^2 \cosh{2r} + 2 \sinh^2{r}$. 

For measurements of the $G(1, 1)$ correlation function, the results are again deferred to the companion Mathematica file. The optimized $\Delta \epsilon_{G(1, 1)}^2$ in the regime of large photon numbers and in the absence of single-photon losses read
\begin{equation}
    \Delta \epsilon_{G(1, 1)}^2 \approx \frac{10}{n_T^3},
\end{equation}
and similarly, we find $\Delta \epsilon^2_{g(1,1)}$ for the normalized correlation function given in the companion Mathematica file, $g(1,1)$. Its lossless optimized version scales as
\begin{equation}
    \Delta \epsilon^2_{g(1,1)} \approx \frac{1}{n_T^2}.
\end{equation}
We compare these results in Fig. \ref{errorVSnTthreeCa}. It is apparent that
$G(1,1)$ outperforms $g(1,1)$ which is superior to the NRF measurement case. 

Next, we investigate the effect of single-photon losses on the estimation error $\Delta \epsilon^2$. To this end, we introduce the normalized error ($\Delta \epsilon^2_{\text{loss}}/\Delta \epsilon^2_{\text{lossless}}$). A ratio of one indicates that the single-photon losses do not affect the $\Delta \epsilon^2$, hence the observed enhancements are robust against experimental imperfections. On the other hand, a ratio larger than one measures the degradation of the achievable precision due to experimental imperfection. 
We present the results in the middle panels of Fig.~\ref{NormalisedError} for different rates of single-photon loss. 

For different correlation function, the normalized errors quickly saturate to a fixed number as a function of $n_T$, indicating that the scaling behavior is not affected by increasing the total photon number incident on the sample and the effects of the single photon loss can not be compensated by increasing the total photon number except for $g(1,1)$. The estimation error $\Delta \epsilon^2$ associated with the measurements of the NRF shows the strongest degradation as a function of the single-photon loss compared to $G(1,1)$ and $g(1,1)$. 
We observe an $\approx 13$-fold degradation of the $\Delta \epsilon_{NRF}^2$ for $\eta=0.7$ compared to the lossless case (see middle panel of Fig. \ref{NormalisedError}). 
In contrast, $G(1,1)$ only shows a small change. The normalized error yielded $\approx 1.17$ for $\eta=0.7$. Contrary to other correlation functions, the optimized $g(1,1)$ is almost independent of single photon loss, approaching the constant $1$ at sufficiently large photon numbers. 
This indicates that in the regime of the large photon number, the effects of single-photon loss in optimized $\Delta \epsilon^2$ of $g(1,1)$ will be completely eroded.  

Next, we discuss the optimal amount of squeezing and seeding to maximize the precision at a fixed photon number. 
The total photon number, $n_T=\alpha_1^2 \cosh{2r} + 2 \sinh^2{r}$, is a combination of a squeezed vacuum contribution ($2 \sinh^2{r}$) and a contribution from the (squeezed) coherent seed ($\alpha_1^2 \cosh{2r}$). Through the optimization, we fix the photon number and adjust the squeezing parameter to obtain the best precision. This indicates that we are tuning the competition between the squeezed vacuum contribution and the coherent seed in the total photon number. We presented these results in Fig. \ref{SSnTrForNRFG11g2}.  
Evidently, in the NRF measurements, the contributions from both seeding and vacuum squeezing total photon number should be identical, i.e. $2 \sinh^2{r} \simeq \alpha_1^2 \cosh{2r}$. For the $G(1,1)$ on the other hand, we need very weak squeezing ($r \approx 0.5$) and very large seeding to optimize $\Delta \epsilon^2$. Finally, $g(1,1)$ is optimized by a fixed, very small amount of seeding ($\alpha_1^2 \cosh{2r} \approx 2$).

\subsection{Squeezed double-seeded coherent state}

Here, we consider seeding the OPA with two coherent states with amplitudes $\alpha_1$ and $\alpha_2$, and phases $\phi_1$ and $\phi_2$, respectively. Similar to the previous section, we optimize the squeezing parameter and relative phase depending on the single-photon loss for fixed total photon numbers at the sample. We present the results of such optimizations in Fig. \ref{SSnTrForNRFG11g2} and Fig. \ref{DSphaseNRFG11g2lossless}.

In contrast to the previous scenarios, here the phases of OPA and coherent states affect the estimation error $\Delta \epsilon^2$. We plot the dependence of the $\Delta \epsilon^2$ on $\theta$ and relative coherent state phase $\Phi$ (= $\phi_1 + \phi_2$) in Fig.~\ref{DSphaseNRFG11g2lossless}. For NRF, $\theta-\Phi =0$ and $2\pi$ give the best  $\Delta \epsilon^2$ (left panel of Fig. \ref{DSphaseNRFG11g2lossless}). This is in contrast to $G(1,1)$ where the optimal choice for the relative phase is $\theta-\Phi = \pm \pi$ (middle panel of Fig. \ref{DSphaseNRFG11g2lossless}). Finally, for $g(1,1)$, we find the optimum phase as $\theta-\Phi \approx 10\pi/11$ and $12\pi/11$ independent from the photon number (right panel of Fig. \ref{DSphaseNRFG11g2lossless}).

The estimation error $\Delta \epsilon^2$ for all three correlation functions are given in the companion Mathematica file. Optimal input states give rise to the following approximate scaling behaviours in the absence of the single-photon loss 
\begin{equation}
    \Delta \epsilon^2_{NRF} \approx \frac{1}{n_T^2},
\end{equation}
\begin{equation}
    \Delta \epsilon^2_{G(1,1)} \approx \frac{6}{n_T^{3.5}},
\end{equation}
and
\begin{equation}
    \Delta \epsilon^2_{g(1,1)} \approx \frac{1}{n_T^2},
\end{equation}
respectively. We are thus able to improve the precision of measuring TPA absorbance for NRF and $G(1,1)$ compared to their squeezed single-seeded counterparts while $g(1,1)$ remains the same (see lower panel of Fig. \ref{errorVSnTthreeCa}). Similar to the squeezed single-seeded measurements, we find that optimized $G(1,1)$ correlation measurements provide the most precise estimates at large photon fluxes.

\begin{table}
\caption{\label{table1} Comparison of the scaling behaviour of the estimation error for different measurement strategies and results for different input states in the absence of the single-photon loss.}
\begin{ruledtabular}
\begin{tabular}{|c|ccc|}
Initial state & $\Delta \epsilon^2_{NRF}$ & $\Delta \epsilon^2_{G(1,1)}$ & $\Delta \epsilon^2_{g(1,1)}$\\ \hline
Two-mode squeezed vacuum & - & $\frac{5}{9 n_T^2}$ & $\frac{1}{n_T^2}$ \\
Single seeding & $\frac{3}{n_T^2}$ & $\frac{10}{n_T^3}$ & $\frac{1}{n_T^2}$\\
Double seeding & $\frac{1}{n_T^2}$ & $\frac{6}{n_T^{3.5} }$ & $\frac{1}{n_T^2}$\\
Double coherent states & $\frac{8}{n_T^2}$ & $\frac{4}{n_T^3}$ & $\frac{4}{n_T^2}$\\
\end{tabular}
\end{ruledtabular}
\end{table}
In order to study the effect of single-photon loss, we plot the right panels of Fig. \ref{NormalisedError} for normalized error. Similar to the single-seeded case, single-photon loss has almost no effects on optimized $\Delta \epsilon^2_{g(1,1)}$, or any effects can be compensated by increasing the total number of the photons. Contrary to the single-seeded case, we observe the largest variation in optimized $\Delta \epsilon^2$ as a function of the single-photon loss for $G(1,1)$. NRF is affected by the single-photon loss and the effects of the single-photon loss can be lowered by increasing the number of the photons reaching the sample. This is not the case for optimized $\Delta \epsilon^2_{G(1,1)}$ where the increment in the number of photons causes the normalized error to deteriorate.

Finally, we compare the strength of the seeding and vacuum squeezing required to reach the optimal $\Delta \epsilon^2$ for the correlation functions with the squeezed double seeding probes. The total mean photon number $n_T$ ($=\alpha_1^2 - \alpha_2^2 + (1 + \alpha_1^2 + \alpha_2^2) \cosh{2r} + 2 \alpha_1 \alpha_2 \cos{\theta} \sinh{2r} + 2 \sinh^2{r}$) is composed of a spontaneous downconversion contribution ($2 \sinh^2{r}$) and photons due to seeded squeezing ($\alpha_1^2 - \alpha_2^2 + (1 + \alpha_1^2 + \alpha_2^2) \cosh{2r} + 2 \alpha_1 \alpha_2 \cos{\theta} \sinh{2r}$). To understand the interrelation between the required squeezing and seeding for the optimal values of $\Delta \epsilon^2$, we plot the variation of the squeezing parameter $r$ with $n_T$ for the three cases in the lower panels of Fig. \ref{SSnTrForNRFG11g2}, showing us that (i) In the case of NRF, we need stronger squeezing for double seeding than the single seeding case. To obtain the optimum $\Delta \epsilon^2$, stronger squeezing than seeding is required for the double seeding input field. (ii) In the case of $G(1,1)$, we need stronger squeezing than the single seeding case. We optimize $\Delta \epsilon^2$ with a high level of seeding and a very low level of squeezing. It should be noted that, for double seeding, the optimum squeezing parameter is an increasing function of the photon number while it is a decreasing function of the photon number for single seeding (compare middle panels of Fig. \ref{SSnTrForNRFG11g2}). (iii) For $g(1,1)$, similar to the single seeding case, we need a very small amount of seeding meaning that $\alpha_1^2 - \alpha_2^2 + (1 + \alpha_1^2 + \alpha_2^2) \cosh{2r} + 2 \alpha_1 \alpha_2 \cos{\theta} \sinh{2r} \approx 2$ would be satisfied for optimal error measurement.

\subsection{Discussions} \label{Discussion}

First, we compare the best precision obtained by squeezing for different measurement schemes to their classical counterparts. We give the results in Fig. \ref{ClassVSQuant}, in which dashed lines correspond to classical transmission measurements with coherent states and the shaded region shows the difference between the corresponding classical and optimised quantum-enhanced measurements. Even though squeezing may always enhance the precision, the achievable gain strongly depends on the correct observable. We obtain the best $\Delta \epsilon^2$ for optimized $G(1,1)$ measurement with squeezed light and find the second-best $\Delta \epsilon^2$ for $G(1,1)$ measurement with the classical light field. This shows that $G(1,1)$ measurements provide better precision in measuring the absorbance of TPA compared to other correlation functions and double seeding compared to vacuum and single seeding. These results with approximate scaling behaviour at large photon fluxes are presented in Table \ref{table1}.

The $g(1,1)$ correlation function is, in practice, a normalized version of the $G(1,1)$. Therefore, we were expecting the induced effects due to the single-photon loss scale identically or at least closely for both the variance of $g(1,1)$ and its modification at the sample (${\partial \langle g(1,1) \rangle}/{\partial \epsilon}$). This is the observation we had in our diagrams (Fig. \ref{NormalisedError}) where the normalized error converged to one for different seeding cases with different degrees of single-photon loss. The optimization procedure presented in Fig. \ref{SSnTrForNRFG11g2} showed that seeding (being single- or double-seeded) and single-photon loss have no effects on the optimization of squeezing parameter for $g(1,1)$ due to $g(1,1)$ being normalized. 
Subsequently, their impacts on scaling behaviour also became irrelevant. Therefore, in $g(1,1)$ measurements, we have the advantage of removing the detrimental impact of single-photon losses on the estimation error ($\Delta \epsilon^2$) but at the cost of forgoing any quantum nature of light and seeding advantages in the scaling behaviour. 

The effects of the type of seeding (single or double) become relevant for $G(1,1)$. While for single seeding, optimization results in requiring a small squeezing parameter for optimum estimation error, we observed the opposite for double seeding (compare middle panels of Fig. \ref{SSnTrForNRFG11g2}). These differences in the behaviours are rooted in the change of the variance and mean photon flux with the sample absorbance (${\partial \langle G(1,1) \rangle}/{\partial \epsilon}$) from single-seeded to double-seeded cases for $G(1,1)$. In single seeding, the variance grows faster as a function of the squeezing parameter than the photon flux at the sample. Therefore, a small value for the squeezing parameter optimizes the sensitivity. The opposite happens for double seeding where the photon flux at the sample increases faster compared to the variance as a function of the squeezing parameter.   
Finally, in $G(1,1)$ measurements, we cannot completely compensate the effects of the single-photon loss in the measurement. 

Lastly, we focus on the NRF. From the formula of the NRF \eqref{eq.NRF}, we find that the numerator of the $\Delta \epsilon^2$ of NRF contains higher orders of the single-photon loss parameter. In addition, given the orders of the expectation values of different quantities in the numerator of NRF, we expect to obtain a rather poor estimation error for this correlation function. This has been confirmed in our study where the estimation error of the NRF was worse than other correlation functions for different scenarios of input light fields.
Similarly, we also found that single-photon loss strongly deteriorates the scaling behaviour of the estimation error. We should mention that while we observed improvement in deterioration by going from single seeding to double seeding, the estimation error of the other correlation functions outperformed NRF. The unique feature of NRF was the ability to switch TPA detection on and off and thus distinguish single- from two-photon losses.

\section{Conclusions} \label{Conclusion}

In this paper, we have analyzed non-degenerate two-photon absorption with classical and squeezed states of light. By optimizing the squeezed state of the light, we showed how to minimize the estimation error of TPA in different measurement setups. 

We obtained the best enhancement for the intensity correlation (compared to other correlation functions) where we reported photon flux with the power of $3.5$ as the scaling behaviour of the estimation error. While the presence of the single-photon loss degraded this scaling behaviour, 
it still features the best sensitivity. In contrast, we showed that the normalized intensity correlation function is robust against experimental imperfection but this comes at the cost of forgoing enhanced error scaling. Finally, for the NRF, we observed an improvement in the estimation error by utilizing the squeezed light field. Despite this improvement, the performance of NRF in estimation error was worse than other correlation functions. The interesting feature of the NRF is its ability to detect the level of single-photon loss, hence experimental imperfection in the setup. While we have studied the transmission measurements in our paper, it will be interesting to investigate whether greater improvements are feasible in nonlinear interferometers. We leave these studies for the future.

\begin{acknowledgments} 
 
GS acknowledges the financial support from the Institution of Eminence (IoE), Banaras Hindu University, Varanasi under the ``International Visiting Student Program'' scheme and the University Grants Commission (UGC) for the UGC Research Fellowship. S. P. acknowledges support from the Hamburg Quantum Computing Initiative (HQIC) project EFRE. The project is co-financed by ERDF of the European Union and by “Fonds of the Hamburg Ministry of Science, Research, Equalities and Districts (BWFGB)”. DKM acknowledges financial support from the Science \& Engineering Research Board (SERB), New Delhi for CRG Grant (CRG/2021/005917) and Incentive Grant under Institution of Eminence (IoE), Banaras Hindu University, Varanasi, India. FS acknowledges support from the Cluster of Excellence `Advanced Imaging of Matter' of the Deutsche Forschungsgemeinschaft (DFG) - EXC 2056 - project ID 390715994.

\end{acknowledgments}

\bibliography{bibliography_photons}

\appendix

\newpage

\begin{widetext}

\section{Error propagation}\label{appendix error}
Let, $\hat{O}$ be a function of $\epsilon$, its Taylor expansion can be written as
\begin{equation}
    \langle \hat{O} \rangle =  \left.\langle \hat{O} \rangle\right|_{\epsilon = 0}  + \left. \frac{\partial \langle \hat{O} \rangle}{\partial \epsilon} \right|_{\epsilon = 0} \epsilon. \label{Taylor}
\end{equation}
Here, we ignore the higher-order terms of $\epsilon$ because $\epsilon \ll 1$. Therefore, variance of the $\hat{O}$ can be written as \cite{taylor1982introduction}
\begin{equation}
    \text{Var}(\hat{O}) = \langle ( \hat{O} - \langle \hat{O} \rangle )^2 \rangle = \left(\frac{\partial \langle \hat{O} \rangle}{\partial \epsilon} \right)^2_{\epsilon = 0} \Delta \epsilon^2. \label{var}
\end{equation}
Where $\hat{O}$ be function of $n$ variables $x_1, x_2,..., x_n$, which in turn depend on $\epsilon$, we can write
\begin{equation}
    \frac{\partial \langle \hat{O} \rangle}{\partial \epsilon} = \frac{\partial \langle \hat{O} \rangle}{\partial x_1} \frac{\partial x_1}{\partial \epsilon} + \frac{\partial \langle \hat{O} \rangle}{\partial x_2} \frac{\partial x_2}{\partial \epsilon} + ... + \frac{\partial \langle \hat{O} \rangle}{\partial x_n} \frac{\partial x_n}{\partial \epsilon}.
\end{equation}
Here $x_i=\langle \hat{x_i} \rangle$ ($i=1,2,...,n$) are expectation value w.r.t. the final state of the system. Again, we can write $\langle \hat{O} \rangle$ in terms of variables $x_1, x_2,..., x_n$ as \cite{taylor1982introduction},
\begin{equation}
    \langle \hat{O} \rangle = \left.\langle \hat{O} \rangle\right|_{\epsilon = 0} + \frac{\partial \langle \hat{O} \rangle}{\partial x_1} \Delta x_1 + \frac{\partial \langle \hat{O} \rangle}{\partial x_2} \Delta x_2 + ... + \frac{\partial \langle \hat{O} \rangle}{\partial x_n} \Delta x_n.
\end{equation}
Where, $\Delta x_i = \sqrt{\hat{x_i} - \langle \hat{x_i} \rangle}$ ($i=1,2,...,n$) denote the stander deviation in the operator. Variance of $\hat{O}$ becomes,
\begin{equation}
    \text{Var}(\hat{O}) = A \mathcal{M} A^T, \label{var2}
\end{equation}
where
\begin{equation}
    A = \begin{bmatrix} \frac{\partial \langle \hat{O} \rangle}{\partial x_1} & \frac{\partial \langle \hat{O} \rangle}{\partial x_2} & \ldots & \frac{\partial \langle \hat{O} \rangle}{\partial x_n}
\end{bmatrix},
\end{equation}
and covariance matrix
\begin{equation}
    \mathcal{M} = \begin{bmatrix}

    \langle (\Delta x_1)^2 \rangle & \langle \Delta x_1  \Delta x_2  \rangle & \ldots & \langle \Delta x_1  \Delta x_n  \rangle \\
    \langle \Delta x_2  \Delta x_1  \rangle & \langle (\Delta x_2)^2 \rangle & \ldots & \langle \Delta x_2  \Delta x_n  \rangle \\
    \vdots & \vdots & \ldots & \vdots \\
    \langle \Delta x_n  \Delta x_1  \rangle & \langle \Delta x_n  \Delta x_2  \rangle & \ldots & \langle (\Delta x_n)^2 \rangle\\
\end{bmatrix}.
\end{equation}
Here, the covariance matrix accounts for the impact of correlations between the operators. If operators are uncorrelated, the off-diagonal elements of the covariance matrix reduce to zero.

From Eq. (\ref{var}) and Eq. (\ref{var2}), we can write
\begin{equation}
    \Delta \epsilon^2 = \frac{ A \mathcal{M} A^T }{ \left( \frac{\partial \langle \hat{O} \rangle}{\partial \epsilon}\right)^2_{\epsilon = 0} }
\end{equation}

\section{Estimation error in NRF}

We can write the Eq. (\ref{eq.NRF}) as
\begin{equation}
    \mathcal{N} = \frac{1}{\langle \hat{n}_1 \rangle + \langle \hat{n}_2 \rangle} \left( \langle \hat{n}_1^2 \rangle + \langle \hat{n}_2^2 \rangle - 2\langle \hat{n}_1 \hat{n}_2 \rangle -\langle \hat{n}_1 \rangle^2 - \langle \hat{n}_2 \rangle^2 + 2 \langle \hat{n}_1 \rangle \langle \hat{n}_2 \rangle \right). \label{A12}
\end{equation}
From Eq. (\ref{Taylor}), we can write 
\begin{equation}
    \mathcal{N} = \left. \mathcal{N} \right|_{\epsilon=0} + \left. \frac{\partial \mathcal{N}}{\partial \epsilon} \right|_{\epsilon=0} \epsilon, \label{AA1}
\end{equation}
where, 
\begin{equation}
    \frac{\partial \mathcal{N}}{\partial \epsilon} = \left( \frac{\partial \mathcal{N}}{\partial \langle \hat{n}_1^2 \rangle}\frac{\partial \langle \hat{n}_1^2 \rangle}{\partial \epsilon} + \frac{\partial \mathcal{N}}{\partial \langle  \hat{n}_2^2 \rangle}\frac{\partial \langle  \hat{n}_2^2 \rangle}{\partial \epsilon} +  \frac{\partial \mathcal{N}}{\partial \langle \hat{n}_1 \hat{n}_2 \rangle}\frac{\partial \langle \hat{n}_1 \hat{n}_2 \rangle}{\partial \epsilon} + \frac{\partial \mathcal{N}}{\partial \langle \hat{n}_1 \rangle}\frac{\partial \langle \hat{n}_1 \rangle}{\partial \epsilon} + \frac{\partial \mathcal{N}}{\partial \langle  \hat{n}_2 \rangle}\frac{\partial \langle  \hat{n}_2 \rangle}{\partial \epsilon} \right). \label{NRFpartial}
\end{equation}
Variance can be written as
\begin{equation}
    \text{Var} (\mathcal{N}) = \left(\frac{\partial \mathcal{N}}{\partial \epsilon}\right)^2_{\epsilon=0} \Delta \epsilon^2. 
\end{equation}
Again, we can write the Eq. (\ref{A12}) as
\begin{equation}
    \mathcal{N} = \left. \mathcal{N} \right|_{\epsilon=0} + \left( \frac{\partial \mathcal{N}}{\partial \langle \hat{n}_1^2 \rangle}{\langle \hat{n}_1^2 \rangle} + \frac{\partial \mathcal{N}}{\partial \langle  \hat{n}_2^2 \rangle}{\langle  \hat{n}_2^2 \rangle}+ \frac{\partial \mathcal{N}}{\partial \langle \hat{n}_1 \hat{n}_2 \rangle}{\langle \hat{n}_1 \hat{n}_2 \rangle} + \frac{\partial \mathcal{N}}{\partial \langle \hat{n}_1 \rangle}{\langle \hat{n}_1 \rangle} + \frac{\partial \mathcal{N}}{\partial \langle  \hat{n}_2 \rangle}{\langle  \hat{n}_2 \rangle} \right). \label{A3}
\end{equation} 
From Eq. (\ref{var2}),
\begin{equation}
    \text{Var} (\mathcal{N}) = 
    A_{\mathcal{N}} \mathcal{M}_{\mathcal{N}} A_{\mathcal{N}}^T, \label{NRFvariance}
\end{equation}
where
\begin{equation}
A_{\mathcal{N}} = \begin{bmatrix}
    \frac{\partial \mathcal{N}}{\partial \langle \hat{n}_1^2 \rangle} & \frac{\partial \mathcal{N}}{\partial \langle  \hat{n}_2^2 \rangle} & \frac{\partial \mathcal{N}}{\partial \langle \hat{n}_1 \hat{n}_2 \rangle} & \frac{\partial \mathcal{N}}{\partial \langle \hat{n}_1 \rangle} & \frac{\partial \mathcal{N}}{\partial \langle  \hat{n}_2 \rangle}
    \end{bmatrix}
\end{equation} and
    
\begin{equation}
    \mathcal{M}_{\mathcal{N}} = \begin{bmatrix}
    \langle (\Delta \hat{n}_1^2)^2 \rangle & \langle (\Delta \hat{n}_1^2) ( \Delta \hat{n}_2^2) \rangle & \langle (\Delta \hat{n}_1^2 \Delta \hat{n}_1 \hat{n}_2) \rangle & \langle (\Delta \hat{n}_1^2 \Delta \hat{n}_1) \rangle & \langle (\Delta \hat{n}_1^2 \Delta \hat{n}_2) \rangle \\
    \langle (\Delta \hat{n}_2^2\Delta \hat{n}_1^2) \rangle & \langle (\Delta \hat{n}_2^2)^2 \rangle & \langle (\Delta \hat{n}_2^2 \Delta \hat{n}_1 \hat{n}_2) \rangle & \langle (\Delta \hat{n}_2^2 \Delta \hat{n}_1) \rangle & \langle (\Delta \hat{n}_2^2 \Delta \hat{n}_2) \rangle \\
    \langle (\Delta \hat{n}_1 \hat{n}_2\Delta \hat{n}_1^2) \rangle & \langle (\Delta \hat{n}_1 \hat{n}_2 \Delta \hat{n}_2^2) \rangle & \langle (\Delta \hat{n}_1 \hat{n}_2)^2 \rangle & \langle (\Delta \hat{n}_1 \hat{n}_2 \Delta \hat{n}_1) \rangle & \langle (\Delta \hat{n}_1 \hat{n}_2 \Delta \hat{n}_2) \rangle \\
    \langle (\Delta \hat{n}_1\Delta \hat{n}_1^2) \rangle & \langle (\Delta \hat{n}_1 \Delta \hat{n}_2^2) \rangle & \langle (\Delta \hat{n}_1 \Delta \hat{n}_1 \hat{n}_2) \rangle & \langle (\Delta \hat{n}_1)^2 \rangle & \langle (\Delta \hat{n}_1 \Delta \hat{n}_2) \rangle\\
    \langle (\Delta \hat{n}_2 \Delta \hat{n}_1^2) \rangle & \langle (\Delta \hat{n}_2 \Delta \hat{n}_2^2) \rangle & \langle (\Delta \hat{n}_2 \Delta \hat{n}_1 \hat{n}_2) \rangle & \langle (\Delta \hat{n}_2 \Delta \hat{n}_1) \rangle & \langle (\Delta \hat{n}_2)^2 \rangle\\
    \end{bmatrix}.
\end{equation}
Off-diagonal elements are considered the correlation between the operators.

\section{Estimation error in \texorpdfstring{$g(1,1)$}{Lg}}

We can write the Eq. (\ref{eq.g2}) as
\begin{equation}
    g(1,1) = \left. g(1,1)\right|_{\epsilon=0} + \left. \frac{\partial g(1,1)}{\partial \epsilon} \right|_{\epsilon=0} \epsilon, \label{A1}
\end{equation}
where, 
\begin{equation}
\frac{\partial g(1,1)}{\partial \epsilon} = \left( \frac{\partial g(1,1)}{\partial \langle \hat{n}_1 \hat{n}_2 \rangle}\frac{\partial \langle \hat{n}_1 \hat{n}_2 \rangle}{\partial \epsilon} + \frac{\partial g(1,1)}{\partial \langle \hat{n}_1 \rangle}\frac{\partial \langle \hat{n}_1 \rangle}{\partial \epsilon} + \frac{\partial g(1,1)}{\partial \langle  \hat{n}_2 \rangle}\frac{\partial \langle  \hat{n}_2 \rangle}{\partial \epsilon} \right).\label{g2partial}    
\end{equation}
From Eq. (\ref{var}), we can write
\begin{equation}
    \text{Var} (g(1,1)) = \left( \frac{\partial g(1,1)}{\partial \epsilon} \right)^2_{\epsilon=0} \Delta \epsilon^2, \label{A2}
\end{equation}
Again, we can write the Eq. (\ref{eq.g2}) as
\begin{equation}
    g(1,1) = \left. g(1,1)\right|_{\epsilon=0} + \left( \frac{\partial g(1,1)}{\partial \langle \hat{n}_1 \hat{n}_2 \rangle}{\langle \hat{n}_1 \hat{n}_2 \rangle} + \frac{\partial g(1,1) (\epsilon)}{\partial \langle \hat{n}_1 \rangle}{\langle \hat{n}_1 \rangle} + \frac{\partial g(1,1) (\epsilon)}{\partial \langle  \hat{n}_2 \rangle}{\langle  \hat{n}_2 \rangle} \right), \label{g2A3}
\end{equation} 
and similar to the G(1,1) case, this gives us
\begin{equation}
    \text{Var} (g(1,1)) = 
    \begin{bmatrix}
    \frac{\partial g(1,1)}{\partial \langle \hat{n}_1 \hat{n}_2 \rangle} & \frac{\partial g(1,1)}{\partial \langle \hat{n}_1 \rangle} & \frac{\partial g(1,1)}{\partial \langle  \hat{n}_2 \rangle}
    \end{bmatrix}
\begin{bmatrix}
    \langle (\Delta \hat{n}_1 \hat{n}_2)^2 \rangle & \langle (\Delta \hat{n}_1 \hat{n}_2 \Delta \hat{n}_1) \rangle & \langle (\Delta \hat{n}_1 \hat{n}_2 \Delta \hat{n}_2) \rangle \\
    \langle ( \Delta \hat{n}_1 \Delta \hat{n}_1 \hat{n}_2) \rangle & \langle (\Delta \hat{n}_1)^2 \rangle & \langle (\Delta \hat{n}_1 \Delta \hat{n}_2) \rangle \\
    \langle ( \Delta \hat{n}_2 \Delta \hat{n}_1 \hat{n}_2) \rangle &  \langle ( \Delta \hat{n}_2 \Delta \hat{n}_1) \rangle &  \langle (\Delta \hat{n}_2)^2 \rangle \\
\end{bmatrix}
\begin{bmatrix}
    \frac{\partial g(1,1)}{\partial \langle \hat{n}_1 \hat{n}_2 \rangle} \\
    \frac{\partial g(1,1)}{\partial \langle \hat{n}_1 \rangle} \\
    \frac{\partial g(1,1)}{\partial \langle  \hat{n}_2 \rangle}
    \end{bmatrix}. \label{g2variance}
\end{equation}

\end{widetext}

\end{document}